\documentclass[aps,prb,floatfixtwocolumn,showpacs,superscriptaddress,amsmath,amssymb,reprint]{revtex4-1}
\usepackage{verbatim}
\usepackage{amsmath}
\usepackage{amssymb}
\usepackage{graphicx}
\usepackage[english,american]{babel}
\usepackage{hyperref}
\hypersetup{
breaklinks=true,
colorlinks=true,
citecolor=blue,
linkcolor=blue,
filecolor=blue,
urlcolor=blue
}

\makeatletter
\PassOptionsToPackage{caption=false}{subfig} 
\makeatother

\begin{document}
%\selectlanguage{american}%

\title{{\normalsize{} Localization transition in one dimension using Wegner flow
equations}}

\author{{\normalsize{}Victor L. Quito}}
\thanks{Authors 
contributed equally}{\normalsize \par}

\affiliation{Instituto de F\'isica Gleb Wataghin, Unicamp, Rua S\'ergio Buarque
de Holanda, 777, CEP 13083-859 Campinas, SP, Brazil}

\affiliation{Institute of Quantum Information and Matter, Dept. of Physics, 
Caltech,
Pasadena, CA 91125}

\author{{\normalsize{}Paraj Titum}}
\thanks{Authors 
contributed equally}{\normalsize \par}

\affiliation{Institute of Quantum Information and Matter, Dept. of Physics, 
Caltech,
Pasadena, CA 91125}

\author{{\normalsize{}David Pekker}}

\affiliation{Department of Physics and Astronomy, University of Pittsburgh}

\affiliation{Institute of Quantum Information and Matter, Dept. of Physics, 
Caltech,
Pasadena, CA 91125}

\author{{\normalsize{}Gil Refael}}

\affiliation{Institute of Quantum Information and Matter, Dept. of Physics, 
Caltech,
Pasadena, CA 91125}

%\selectlanguage{english}%

\date{{\normalsize{}\today}}
%\selectlanguage{american}%
\begin{abstract}
{\normalsize{}The flow equation method was proposed by Wegner as a
technique for studying interacting systems in one dimension. Here,
we apply this method to a disordered one dimensional model with power-law decaying hoppings. This model presents a transition as function of the decaying exponent
$\alpha$. We derive the flow equations, and the evolution of single-particle operators. The flow equation reveals the delocalized nature of the states for $\alpha<1/2$. Additionally, in the regime, $\alpha>1/2$, we present a strong-bond renormalization group structure based on iterating the three-site clusters, where we solve the flow equations perturbatively. This renormalization group approach allows us to probe the critical point $\left(\alpha=1\right)$. This method correctly reproduces the critical level-spacing statistics, and the fractal dimensionality of the eigenfunctions.}
\end{abstract}

\pacs{71.30.+h, 64.60.ae, 71.23.An, 72.20.Ee, 72.15.Rn}

\maketitle
%\selectlanguage{english}%

\section{{\normalsize{}Introduction \label{sec:Introduction}}}

The interplay of disorder and quantum fluctuation leads to ubiquitous
effects, with the so-called Anderson \cite{AndersonPRL} localization
being one of the most striking. Localization effects emerge from 
quantum interference of the wave function in 
sites randomly displaced in a lattice. Equivalently,
the same effect appears in ordered lattices where the chemical potential
is random. The consequences of Anderson localization have been studied
experimentally and numerically over the past several 
decades \cite{EversRMP,KramerMackinnon1993}.  A scaling analysis 
\cite{GangofFour1979} showed that the typical wave function
in one or two dimensions decays exponentially in a 
non-interacting system with short-range hopping and random chemical potential. Three- and higher- dimensional systems, 
however, possess a delocalization transition, exhibiting 
multifractal wave functions at the critical energy \cite{EversRMP,GruzbergPRL2011}.

Interestingly, a metal-to-insulator transition is also achieved in one-dimension systems when the hoppings are allowed to be long-ranged\cite{LevitovPRL,MirlinPRE,EversMirlinPRL2000}. In this case, the effective system dimension changes with the power-law decaying exponent of the hopping. A localization transition is observed in the states by tuning the power-law decay exponent only, as long as the chemical potential is random. This transition occurs for states at all energies unlike the Anderson transition in short-ranged systems where there is a  mobility edge. Additionally, at the critical point, the full spectrum is characterized by multifractal behavior of the wave functions \cite{MirlinPRE,EversMirlinPRL2000}. 

Localization effects took center-stage again recently, with theory, numerics,
and experiments in cold atoms probing weakly interacting disordered systems 
\cite{BAA2006,Bardarson2012,pekkersdrgxprx,HuseNandkishorePhenom2014,
VoskMBLT2015,BlochMBL2015,Luitz2015,RademakerPRL,Vasseurmbl2015}. The focus of these studies is 
the many-body localized (MBL) state, where electron-electron interactions fail to 
thermalize the system, and the rules of statistical mechanics do not hold 
\cite{HuseNandkishorePhenom2014,VoskMBLT2015,PalHuse2010}. This state is 
predicted to exist even at infinite temperature where
the analyses of highly excited states become relevant \cite{OganesyanHusePRB,monthus2016flow}. For strong interactions, the MBL state undergoes a transition to an ergodic state. Across this transition, the distribution of level-spacing statistics of the full spectrum changes\cite{SerbynMoore2016,Kravtsovloctrans,monthus2016many,monthusentan,monthuslvlrep,Shklovskii1993,Avishai2002,Lea2004}.
This implies a need to develop analytical tools that address the full spectrum
of the Hamiltonian. 

The daunting task of accounting for the behavior of excited states anywhere in 
the spectrum requires a scheme that extracts the important elements in the 
Hilbert space and the Hamiltonian. Such a task has been successfully 
accomplished, for instance, with the SDRG-X technique \cite{pekkersdrgxprx}, a 
generalization of the
Ma and Dasgupta's proposal \cite{Ma1979,DasguptaMa1980}, and recently applied to 
a variety of disordered systems \cite{Martin2015, XuSBRG2016}. Another path to 
such a scheme could be the flow equation technique. This technique was 
introduced by Wegner \cite{wegnerann}, in the context of condensed matter, and, 
concomitantly, by Glasek and Wilson \cite{glasekflow1,glasekflow2}, in the high 
energy physics. Our focus is employing this technique to address localization 
transitions.

In this paper we describe the adaptation of the flow-equation technique to 
study localization transitions in non-interacting one-dimensional systems with long-range hoppings. In particular, we consider hopping 
terms with a random magnitude, and a variance that decays
as a power law with distance. The metal-to-insulator transition
is obtained by tuning the power-law exponent, $\alpha$ (see Fig.
\ref{fig:Phase-diagram}), with the critical point at $\alpha=1$.
The connectivity of the system makes it behave as effectively higher
dimensional, with the dimension related to the power-law exponent
$\alpha$.

The flow analysis we develop allows us to study the full phase diagram of
the power-law hopping non-interacting system. We show that, for $\alpha<0.5$, the distribution
of hoppings flows to an attractive fixed point at $\alpha=0$. This
means that the phase for $\alpha<0.5$ is in the Gaussian orthogonal
ensemble (GOE) with extended states. For $0.5<\alpha<1$, the
states have critical and intermediate statistics. In this regime,
we recast the flow as a controlled strong-bond renormalization group (RG) procedure, and recover the 
full single-particle spectrum with appropriate level statistics. The strong-bond RG flow 
produces the spectrum of energy differences from the largest to smallest, 
iteratively, while also generating a diffusion in the space of hopping 
strengths. The level repulsion for $\alpha<1$ emerges as a consequence of a 
crossover of the hopping distribution function from  power-law to uniform at the 
average level spacing scale. The method is even more successful for $\alpha>1$, 
where localization emerges, associated with Poisson statistics of the level 
spacings.

The flow equation approach and, in particular, the strong-bond RG scheme, 
provides a new and natural framework with which to address localization and 
level statistics in disordered systems. In our presentation we will emphasize 
the universal aspects of the method, as well as its intuitive features. It is 
natural to expect that it could be used in more complicated settings.

This paper is organized as follows.  In Section~\ref{sec:The-model-PBRM},
we review the model of non-interacting particles with power-law hopping, the power-law banded random matrix (PBRM). We briefly explain
the phases that have been previously found by Mirlin \emph{et 
al}.\cite{MirlinPRE}
and Levitov \cite{Levitov2,LevitovPRL}. In 
Section~\ref{sec:Method:-Wegners-Flow},
we introduce the flow equation (FE) method, focusing on its application
to this model at $\alpha<0.5$. The flow reveals an
attractive fixed point at $\alpha=0$. In Section~\ref{sec:Flow-RG}, we introduce 
the
strong-bond RG scheme that consists of eliminating hopping in bonds (as opposed 
to sites, as proposed in Ref. \onlinecite{JavanmardPRB}). We discuss the bond 
selection and how it can be derived from the two-site and three-site flow equations. 
We explain
the appearance of level repulsion as a function of the exponent, 
$\frac{1}{2}<\alpha<1$.

\section{{\normalsize{}The model: PBRM \label{sec:The-model-PBRM}}}

The system we seek to analyze consists of a one-dimensional chain of 
non-interacting particles with random on-site disorder and random hoppings whose 
typical strength decays algebraically with site distance. This is the so-called 
PBRM model. It exhibits an Anderson transition despite 
its low dimensionality. The Hamiltonian
in second-quantized notation is
\begin{equation}
H=\sum_{i,j}J_{i}^{j}c_{i}^{\dagger}c_{j}+\sum_{i}h_{i}c_{i}^{\dagger}c_{i},
\label{eq:PBRMHamiltonian}
\end{equation}
where $h_{i}$ and $J_{i}^{j}$ are random uncorrelated variables.
 The standard deviation of $J_{j}^{i}$ decays with distance as 
$\sigma_{J_{i}^{j}}=\frac{\sigma_{J_{0}}}{\left|i-j\right|^{\alpha}}$.  
No further assumptions regarding the distributions are made at this point, as the phase diagram is independent of the ratio $\frac{\sigma_{h}}{\sigma_{J_{0}}}$, where $\sigma_{h}>0$ is the standard deviation of the $h$ distribution. The operators 
$c_{i}^{\dagger}$ $\left(c_{i}\right)$
creates (annihilates) a particle at site $i$. 

\begin{figure}
\includegraphics[width=1\linewidth]{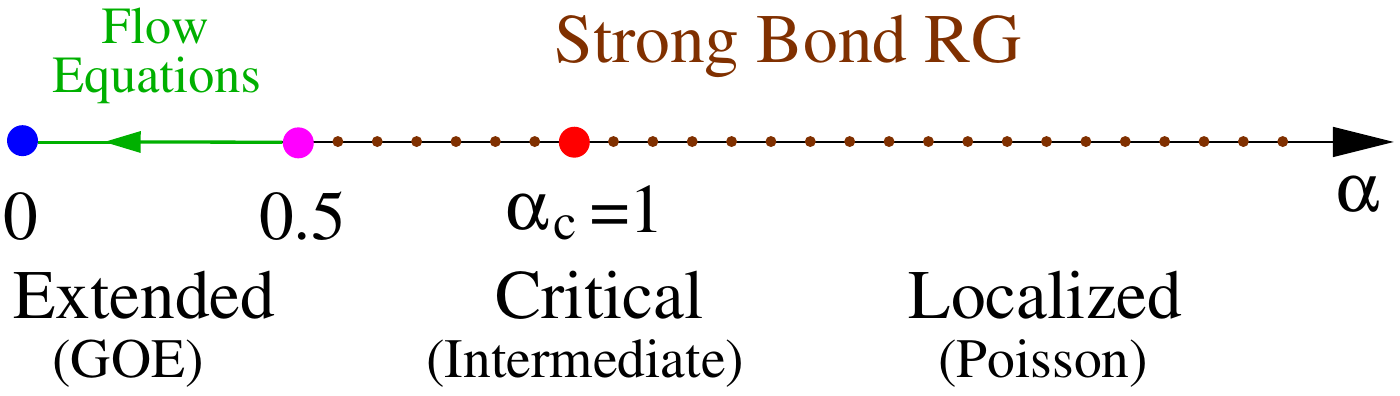}

\caption{Phase diagram of PBRM model, Eq.~(\ref{eq:PBRMHamiltonian}), with 
disordered on-site
potential and random hoppings whose typical value decay with range as a power 
law, $J_{ij}\sim\frac{1}{|i-j|^{\alpha}}$. For $\alpha<\frac{1}{2}$,
the system is equivalent to the $\alpha=0$ Gaussian Orthogonal Ensemble
(GOE). This region is studied in this work via the flow equation technique. A 
strong-bond RG flow scheme based on the flow equations allows us to characterize 
the $\alpha>1/2$ phases. This novel RG scheme we propose does not eliminate any 
degrees of freedom, but consists of a sequence of unitaries. The critical point 
for
the transition to a localized phase is at $\alpha_{c}=1$. The level-spacing
statistics in this phase transitions to Poisson statistics.  
\label{fig:Phase-diagram}}

\end{figure}

The exponent $\alpha>0$, which describes power-law decay of long-range hopping, 
is the \textit{only} tuning parameter for a localization-delocalization
transition (see Fig.~\ref{fig:Phase-diagram}). This model has been previously 
studied both by numerical techniques, such as exact diagonalization 
\cite{EversMirlinPRL2000}, and analytical techniques, such as super-symmetric 
methods \cite{MirlinPRE}
and real-space RG\cite{Levitov2,LevitovPRL,EversMirlinPRB} . In the
following, before proposing a new method to tackle the problem, we
review some of the known properties of the model and give a qualitative
description of the phase transition.

\emph{The localized and delocalized phases}:\emph{ }Let us examine
the model, defined in Eq.~(\ref{eq:PBRMHamiltonian}) for the two
limiting cases, $\alpha=0$ and $\alpha\rightarrow\infty$. In the
limit $\alpha=0$ , the Hamiltonian corresponds to a random matrix
in the Gaussian Orthogonal Ensemble (GOE). The properties of the eigenstates
are given by Random Matrix Theory (RMT). The eigenvalues experience
level repulsion and the level spacing distributions obey the Wigner-Dyson
statistics \cite{mehtaRMT}. The phase is, therefore, delocalized
and all the single-particle orbitals are extended. In the opposite
limit, $\alpha\rightarrow\infty$, only nearest-neighbor interactions
are non-zero and the Hamiltonian realizes an Anderson Insulator 
phase.
In such a phase, all the orbitals are known to be localized~\cite{AndersonPRL}. In contrast
with the delocalized phase, the single-particle energies
are uncorrelated and the level spacing exhibits Poisson statistics
\cite{BeenakkerRMP,mehtaRMT}. 

\emph{The Critical point}: This model exhibits a critical point at
the exponent $\alpha=1$. The eigenstates exhibit multifractality,
and the eigenvalues experience level repulsion with intermediate statistics.

The localization-delocalization transition is driven by the proliferation
of resonant sites at \textit{arbitrarily}
long length scales. Here, we say that two sites $i$ and $j$ are in resonance
when the parameters $J_{i}^{j}$, $h_{i}$ and $h_{j}$ satisfy 
$J_{i}^{j}>\left|h_{i}-h_{j}\right|$.
\foreignlanguage{american}{ Let the probability
of a site in resonance with a site $i$, at a distance $R$, be $P\left(R\right)$.
Assuming a constant density of states $n$, the characteristic level
spacing in a shell of width $dR$ is $\Delta\sim\frac{1}{n\ dR}$,
while the hopping strength is $J\sim\frac{1}{R^{\alpha}}$.
Therefore, the number of resonances between $R$ and $R+dR$ is $P\left(R\right)\ 
dR\propto\frac{J}{\Delta}\sim\frac{1}{R^{\alpha}}\ dR$.
Now, the total number of sites in resonance at any length larger than
$R$ is,}

\selectlanguage{american}%
\[
N_{{\rm res}}=\int_{R}^{N}dR'\ P(R')\sim\begin{cases}
\log\left(\frac{N}{R}\right) & ,\text{ for }\alpha=1\\
\frac{1}{R^{\alpha-1}} & ,\text{ for }\alpha>1\\
N^{1-\alpha} & ,\text{ for }\alpha<1,
\end{cases}
\]
where we keep terms at leading order in system size $N$. We conclude that
in the delocalized phase $\left(\alpha<1\right)$ the number
of resonances diverge and, conversely, in the localized phase 
$\left(\alpha>1\right)$ the
number of resonances does not scale with system size, and, hence,
is negligible at the thermodynamic limit. At the critical point $\alpha=1$ 
$N_{{\rm res}}$ diverges logarithmically, which suggests
a phase transition. A more careful derivation of the above result,
along with the real-space renormalization group scheme at the critical
point have been derived by Levitov\cite{Levitov2,LevitovPRL}. For
completeness, we present a short review of Levitov's method in 
Appendix~\ref{sec:Appendix-Levitovs-analysis}. 

\selectlanguage{english}%

\section{{\normalsize{}Disordered Wegner's flow equations 
\label{sec:Method:-Wegners-Flow}}}

\selectlanguage{american}%
The Flow Equation Technique (FET) was first introduced by Wegner, Glasek and 
Wilson \cite{wegnerann,glasekflow1,glasekflow2}. It iteratively constructs a 
unitary transformation
that continuously diagonalizes a Hamiltonian as a function of some
flow ``time'' $\Gamma$. For a simple example illustrating how to compute the 
flow equations, see Appendix \ref{sec:Appendix-Simple-Example-Spin-1/2}. Going 
back to the model we previously introduced
in Eq.~(\ref{eq:PBRMHamiltonian}), we set the coupling constants
to be functions of $\Gamma$ and split it into two parts, 
$H_{0}\left(\Gamma\right)$
and $V\left(\Gamma\right)$:
\begin{eqnarray}
H\left(\Gamma\right) & = & 
\sum_{i}h_{i}\left(\Gamma\right)c_{i}^{\dagger}c_{i}+\sum_{i,j}J_{i}^{j}
\left(\Gamma\right)c_{i}^{\dagger}c_{j},\label{eq:HamiltonianRGtime}\\
 & = & H_{0}\left(\Gamma\right)+V\left(\Gamma\right)\label{eq:Hamiltoniansplit}
\end{eqnarray}
We also require that the $\Gamma$-dependent Hamiltonian defined in
Eq (\ref{eq:HamiltonianRGtime}) satisfies $H\left(\Gamma=0\right)=H$
(see Eq.~(\ref{eq:PBRMHamiltonian})) and that 
$H\left(\Gamma\rightarrow\infty\right)$
becomes diagonal. In order to obtain the infinitesimal rotation generator,
the Hamiltonian is split into diagonal and off-diagonal parts. Note
that the choice of terms as diagonal and off-diagonal depends on
the choice of basis. We work in the number basis such that $c^{\dagger}c$
is diagonal. Now, following Wegner\cite{wegnerann}, the canonical
generator for the infinitesimal unitary transformations is defined
as
\begin{equation}
\eta\left(\Gamma\right)=\left[H_{0}\left(\Gamma\right),
V\left(\Gamma\right)\right].\label{eq:flow-generator}
\end{equation}
The Hamiltonian flows under the operation of the generator, $\eta$, which is 
expressed through a Heisenberg equation of motion with respect to RG time,
\begin{equation}
\frac{d}{d\Gamma}H\left(\Gamma\right)=\left[\eta\left(\Gamma\right),
H\left(\Gamma\right)\right].\label{eq:HamiltonianRGTimeevolution}
\end{equation}
The unitary operator that diagonalizes the Hamiltonian is 
$U\left(\Gamma\right)=\mathcal{T}_{\Gamma}\exp\left(\int^{\Gamma}
d\Gamma'\eta\left(\Gamma'\right)\right)$,
where $\mathcal{T}_{\Gamma}$ denotes RG-time ordering. This generator ensures 
convergence to a diagonal Hamiltonian in the
limit $\Gamma\rightarrow\infty$ if the condition 
$\mbox{Tr}\left(\frac{dH_{0}}{d\Gamma}V\right)=0$
is fulfilled. This condition is obviously true in the system explored
in this paper, since fermionic (bosonic) operators anticommute (commute).
By using the condition $\mbox{Tr}\left(\frac{dH_{0}}{d\Gamma}V\right)=0$,
it becomes simple to prove that \cite{Kehreinbook}
\selectlanguage{english}%
\begin{equation}
\frac{d}{d\Gamma}\mbox{Tr}\left[V\left(\Gamma\right)\right]^{2}=-2\mbox{Tr}
\left(\eta^{\dagger}\eta\right)\le0,
\end{equation}
and, consequently $V\left(\Gamma\right)=0$ as $\Gamma\rightarrow\infty$.

\selectlanguage{american}%
\begin{figure}
\includegraphics[width=1\linewidth]{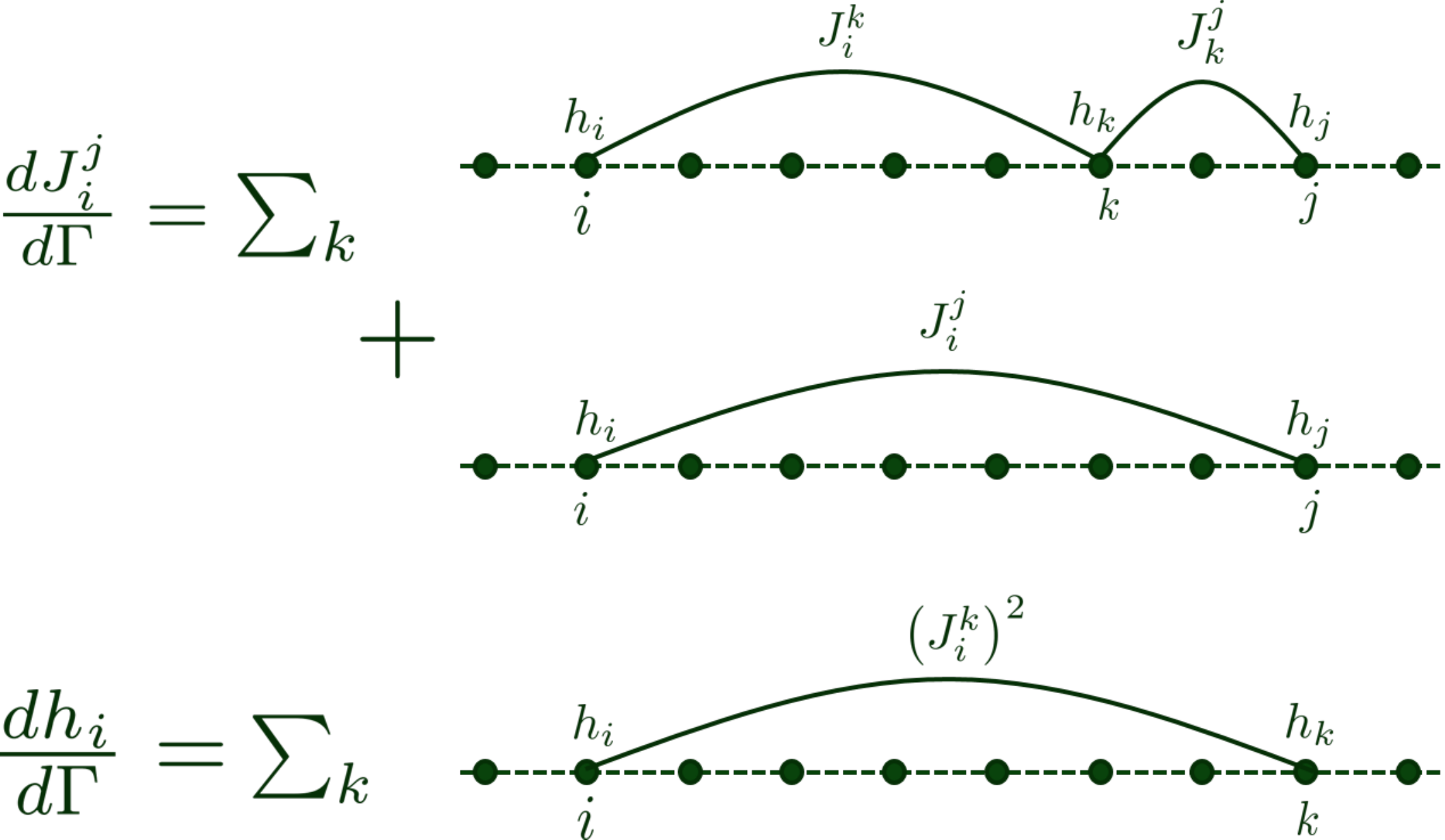}

\raggedright{}\caption{Pictorial representation of the flow equations for the 
hoppings and
fields as calculated in Eqs.~(\ref{eq:J_evol}) and (\ref{eq:h_evol}).
All the contributions are product of three coupling constants. For
the hoppings, the first contribution comes from a sum of terms of
the type $JJh$, that is the product of two hoppings and one field,
while the second contribution comes from $Jhh$, the product of two
fields and one hopping. For the renormalization of hoppings, all contributions
are of type $JJh$. \label{fig:Schematic-flow-equations} }
\end{figure}

\selectlanguage{english}%
The equation of motion obtained in Eq.~(\ref{eq:HamiltonianRGTimeevolution})
leads to the following flow equations for the couplings,%
\begin{comment}
\textbf{{[}Paraj, ok. I added a sentence about it below. It is important
to change all the equations with this renormalization. I have a question
about it... suppose the equation is like}

\begin{eqnarray}
\frac{dJ_{i}^{j}}{d\Gamma}\left(\Gamma\right) & = & 
-4J_{i}^{j}\left(\Gamma\right)\\
\implies\frac{dJ_{i}^{j}}{d\tilde{\Gamma}}\left(\frac{\tilde{\Gamma}}{4}\right) 
& = & -4J_{i}^{j}\left(\frac{\tilde{\Gamma}}{4}\right)
\end{eqnarray}

\textbf{What about the fact that the functions are now function of
$\frac{\tilde{\Gamma}}{4}$? Another approach would be 
$J_{i}^{j}\rightarrow\frac{1}{2}J_{i}^{j}$
and $h_{i}\rightarrow\frac{1}{2}h_{i}$. Once we fix this, I change
the plots{]}}
\end{comment}

\selectlanguage{american}%
\begin{eqnarray}
\frac{dJ_{i}^{j}}{d\Gamma} & = & 
-J_{i}^{j}\left(x_{j}^{i}\right)^{2}-\sum_{k=1}^{N}J_{i}^{k}J_{k}^{j}\left(x_{k}
^{j}-x_{i}^{k}\right),\label{eq:J_evol}\\
\frac{dh_{i}}{d\Gamma} & = & 
-2\sum_{k=1}^{N}\left(J_{k}^{i}\right)^{2}x_{i}^{k},\label{eq:h_evol}
\end{eqnarray}
where we have defined, $x_{j}^{i}=h_{i}-h_{j}$. For convenience,
we have absorbed a factor of $4$ in the definition of $\Gamma$. The initial 
conditions for the couplings are
$J_{i}^{j}\left(\Gamma=0\right)=J_{i}^{j}$ and 
$h_{i}\left(\Gamma=0\right)=h_{i}$.
As a consequence of the Hamiltonian becoming diagonal in the limit
$\Gamma\rightarrow\infty$, we have 
$J_{i}^{j}\left(\Gamma\rightarrow\infty\right)=0$.
The single-particle energy spectrum of the Hamiltonian is obtained
from the set of fields in the end of the flow $\left\{ 
h_{i}\left(\Gamma\rightarrow\infty\right)\right\} $.
The many-body energies can be found by filling these levels. The flow equations 
are represented schematically in Fig.~\ref{fig:Schematic-flow-equations}.

\begin{figure}
\includegraphics[scale=0.4]{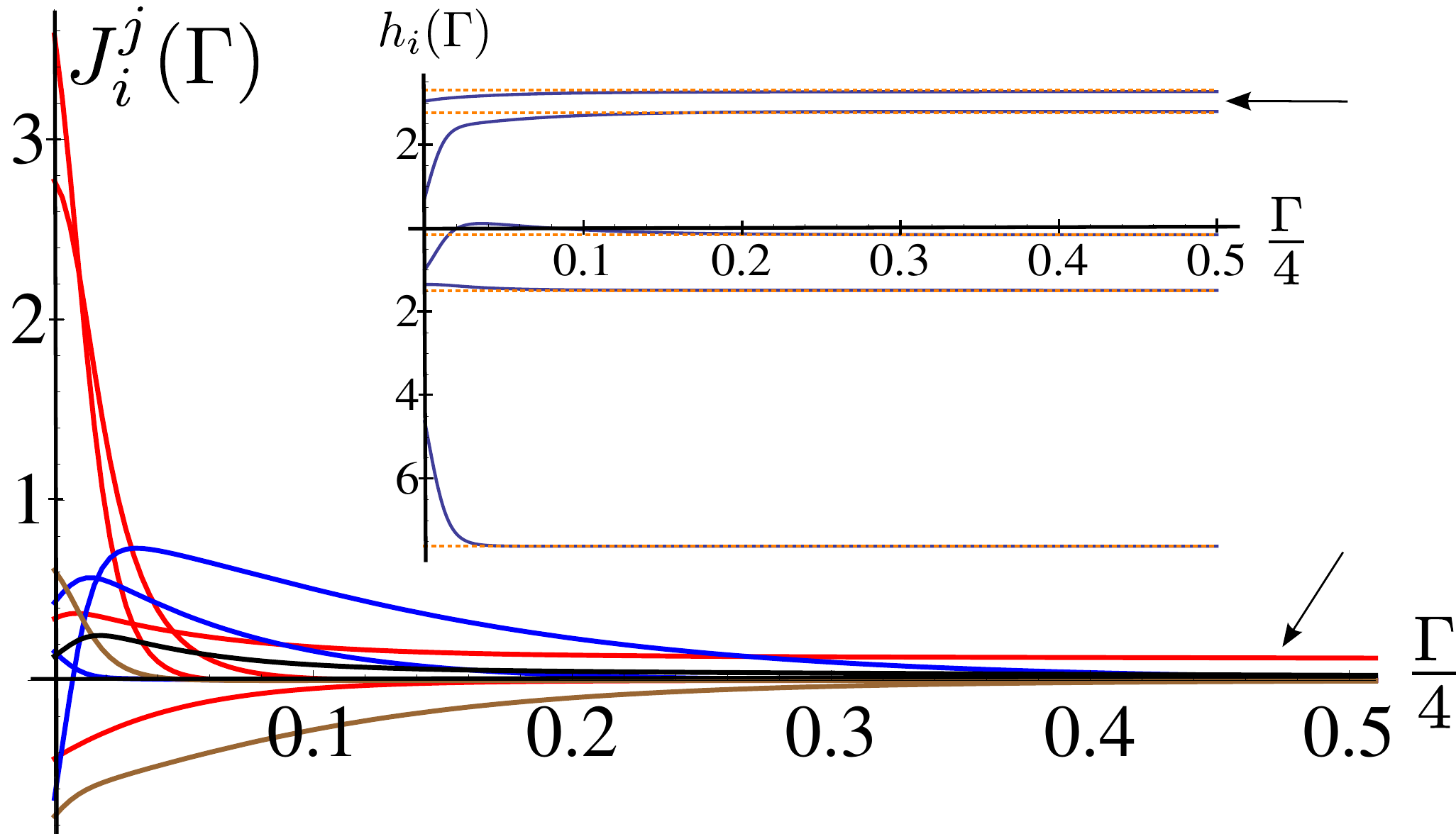}\caption{(Color online) Typical flow for 
the 5-site problem. The initial fields and hoppings
are random variables. The distribution of hoppings is Gaussian, with
a power-law decay with distance $\left|i-j\right|^\alpha$,
$\alpha=1$. The distinct colors represent the different distances
$\left|i-j\right|$ (red, blue, brown, and black curves, in order of increasing 
distance).
Notice that one of the red curves, indicated by the arrow, flows more
slowly to zero. This is due to the fact that the decay term in the $J$ flow is 
proportional to difference of the fields of the two sites connected by it {[}see 
Eq.~(\ref{eq:J_evol}) and the arrow
in the inset curve{]}.  Also shown in the inset is flow of fields (blue)
and their asymptotic approach to the Hamiltonian eigenvalues (horizontal dashed 
orange lines). \label{fig:comparison-flow-exact}}
\end{figure}

The flow equations can be solved numerically, by starting a chain with random 
couplings and evolving them numerically via Eqs.~(\ref{eq:J_evol}) and 
(\ref{eq:h_evol}). In Fig.~\ref{fig:comparison-flow-exact},
we give a comparison of the spectrum obtained using the FE with exact 
diagonalization for a 5 site chain. The decay of $J_{i}^{j}$
is controlled by the field difference, $h_{i}-h_{j}$. When the final
values of $h_{i}$ and $h_{j}$ are close, the decay is much slower,
as can be seen also be seen in the Figure.

In section \ref{sub:Two-site-sol}, we start constructing the phase diagram by exactly solving a chain of two sites. This solution  lends a time scale, that allows for a bond-decimation hierarchy. This forms the foundation for an RG scheme described in Section \ref{sec:Flow-RG}, appropriate for $\alpha>1/2$.  In section \ref{sub:N-site-scaling}, we develop a scaling approach to follow the distributions of bonds under the evolution of the $N$-site flow equations. The  power law exponent of the coupling distribution changes as the couplings flow. From any initial distribution with $\alpha<0.5$, the exponent reaches the universal $\alpha=0$ fixed point. Notice that the combination of the two techniques mentioned above, the direct implementation of the flow equations for $\alpha<1/2$ and the RG scheme developed for $\alpha>1/2$, allows us to map the full phase diagram.

\subsection{Building block: Two-site solution \label{sub:Two-site-sol}}

As a first step, let us solve the illustrative example of the two-site
chain, with fields $h_{1}$ and $h_{2}$ and inter-site hopping, $J\equiv 
J_{1}^{2}$.
It becomes convenient to define a new variable, $x=h_{2}-h_{1}$.
The flow equations, Eqs.~(\ref{eq:J_evol}) and (\ref{eq:h_evol}),
reduce to,
\selectlanguage{american}%
\begin{eqnarray}
\frac{d}{d\Gamma}J\left(\Gamma\right) & = & 
-J\left(\Gamma\right)\left(x\left(\Gamma\right)\right)^{2},\label{
eq:two-site-J_evol}\\
\frac{d}{d\Gamma}x\left(\Gamma\right) & = & 
4\left(J\left(\Gamma\right)\right)^{2}x\left(\Gamma\right).\label{
eq:two-site-h_evol1}
\end{eqnarray}
These equations have a conserved quantity, which we denote
as 
\[
r^{2}=4J\left(\Gamma\right)^{2}+x\left(\Gamma\right)^{2}.
\]
Defining polar coordinates, 
$J\left(\Gamma\right)=\frac{r}{2}\sin\theta\left(\Gamma\right)$
and $x\left(\Gamma\right)=r\cos\theta\left(\Gamma\right)$, we obtain
the flow for $\theta\left(\Gamma\right)$:
\begin{equation}
\frac{d\theta}{d\Gamma}=-\frac{1}{2}r^{2}\sin2\theta\left(\Gamma\right),\label{
eq:two-site-theta_evol}
\end{equation}
where the initial condition is 
$\theta_{0}=\theta\left(0\right)=\arctan\left(\frac{2J}{x}\right)$.
The solution of this equation is 
\begin{equation}
\tan\theta\left(\Gamma\right)=\tan\theta_{0}\exp\left(-r^{2}\Gamma\right).
\end{equation}
Asymptotically, as $\Gamma\rightarrow\infty$, $\theta$ tends to zero: 
$\theta\left(\Gamma\rightarrow\infty\right)=0$. The decay rate of 
$\tan\theta\left(\Gamma\right)$,
gives us a natural RG time scale to achieve a nearly diagonal Hamiltonian: 
\begin{equation}
\tau_{\Gamma}\sim\frac{1}{r^{2}}.
\end{equation}
In this chain of two sites, the master equation for the
distribution of couplings, $J\left(\Gamma\right)$ and $x\left(\Gamma\right)$,
can also be exactly solved. The solution reveals that the distributions
of $\log J\left(\Gamma\right)$ and $x\left(\Gamma\right)$ are correlated, what 
can be tracked back to the constraint that 
$x^{2}\Gamma=-\log\left(J\right)$. Analogous correlations between
$J$ and $x$ variables are also observed for the couplings in larger
chains. The details are provided in 
Appendix~\ref{sub:Appendix-Two-site-solution}.

It is important to note that the two-site flow gives rise to the following 
canonical transformation of the second-quantized creation operators:

\begin{equation}
\left(\begin{array}{c}
\tilde{c}_{1}\\
\tilde{c}_{2}
\end{array}\right)=\left(\begin{array}{cc}
\cos\alpha_{12} & \sin\alpha_{12}\\
-\sin\alpha_{12} & \cos\alpha_{12}
\end{array}\right)\left(\begin{array}{c}
c_{1}\\
c_{2}
\end{array}\right)
\label{2-site-T}
\end{equation}

where $\alpha_{12}=\mbox{sgn}\left(Jx\right)\frac{\theta_{0}}{2}$.

\selectlanguage{english}%

\subsection{$N$-site problem}\label{sub:N-site-scaling}

Now we consider the full coupled flow equations
for the $N$-site problem. Let us start by defining new hopping variables, 
$G_{i}^{j}=J_{i}^{j}l^{-\alpha}$,
where $l=\left|i-j\right|$. We consider the initial distributions for the 
couplings
$J\left(l=\left|j-i\right|\right)$ to have a variance that scales with
length as $\sigma_{J}^{2}\left(l\right)\sim l^{-2\alpha}$, while the $G\equiv 
G_{i}^{j}$ distributions are distance independent. Without loss of generality,
assume that $j>i$. The FE in Eq.~(\ref{eq:J_evol}) rewritten in
terms of $G$ is 
\begin{eqnarray}
-\frac{dG}{d\Gamma} & = & 
\sum_{k=1}^{N}X_{k}\left[\frac{l}{\left|k-i\right|\left|j-k\right|}\right]^{
\alpha}+G\left(x_{j}^{i}\right)^{2},\label{eq:flow-eq-G}\\
 & = & \Delta\left(l\right)+G\left(x\left(l\right)\right)^{2}
\end{eqnarray}
where $X_{k}=G_{i}^{k}G_{k}^{j}\left(x_{k}^{j}-x_{i}^{k}\right)$.
There are two terms in Eq.~(\ref{eq:flow-eq-G}). The term 
$G\left(x\left(l\right)\right)^{2}$
is responsible for the decay in the magnitude of $G$, and $\Delta\left(l\right)$
acts as a random-source term that generates hoppings distributions with changing 
power laws,  which modifies
the distribution of $G$. In order to unveil how this process happens, we ignore 
the decay term for a moment and consider the
scaling of the variance of the distribution
of $\Delta\left(l\right)$ at long distances, $\sigma_{\Delta}\left(l\right)$. 
Let us assume that $X_{k}$ is a scale-independent
uncorrelated random variable, $\left\langle X_{k}X_{k'}\right\rangle 
=\left\langle X^{2}\right\rangle \delta_{kk'}$. With this assumption, we end up 
with
\begin{widetext}
\begin{eqnarray}
\sigma_{\Delta}^{2}\left(l\right) & = & \left\langle X^{2}\right\rangle \left[\sum_{\substack{k=1\\
k\neq\left\{ i,i+l\right\} 
}
}^{N}\frac{l^{2\alpha}}{\left|k-i\right|^{2\alpha}\left|l-\left(k-i\right)\right|^{2\alpha}}\right],\nonumber \\
 & = & \left\langle X^{2}\right\rangle l^{2\alpha}\left[\int_{1}^{l}\frac{dx}{x^{2\alpha}\left(l-x\right)^{2\alpha}} +  \left(\int_{1}^{i-1}+\int_{1}^{N-j}\right)\frac{dx}{x^{2\alpha}\left(l+x\right)^{2\alpha}}\right]
\end{eqnarray}
\end{widetext}
The integral is dominated by possible divergences at
$x=0$ and $x=l$. Consider first  $\alpha<\frac{1}{2}$. It is clear
that we can completely scale out $l$, 
\[
\int_{1}^{l}\frac{dx}{x^{2\alpha}\left(l-x\right)^{2\alpha}}\sim 
l^{1-4\alpha}\int_{\frac{1}{l}}^{1}\frac{dx}{x^{2\alpha}\left(1-x\right)^{
2\alpha}}\propto l^{1-4\alpha},
\]
and, therefore, we expect $\sigma_{\Delta}^{2}\left(l\right)\propto 
l^{1-2\alpha}$.
In contrast, at $\alpha=\frac{1}{2}$, the variance is logarithmically
dependent on $l$, $\sigma_{\Delta}^{2}\left(l\right)\propto\log l$,
hinting a critical behavior. Finally, we note that for the
case of $\alpha>\frac{1}{2}$, the variance is independent of the
length scale, $\sigma_{\Delta}^{2}\left(l\right)\sim\text{const}$.

\begin{figure}
\includegraphics[width=1\linewidth]{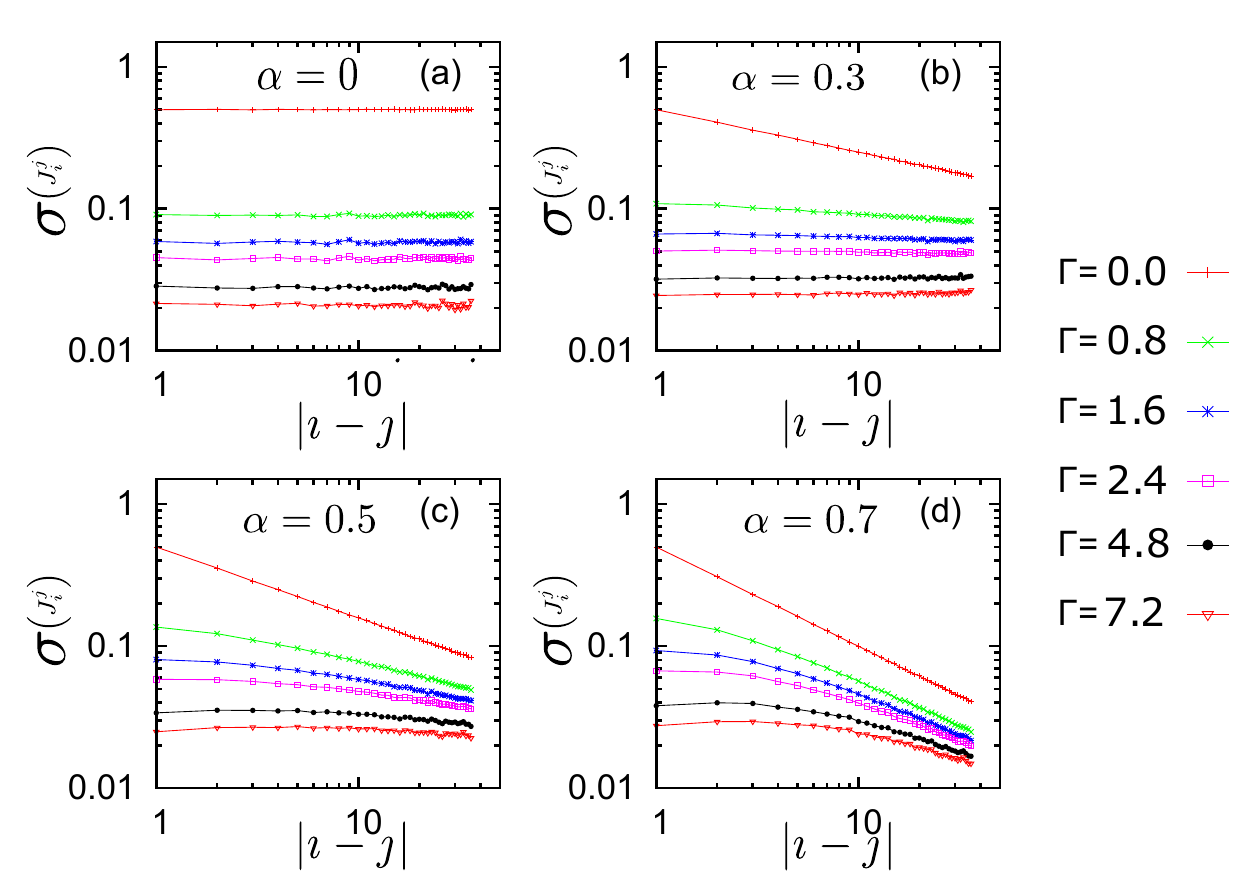}

\caption{(Color Online) The standard deviation of distributions of $J_i^j$, 
$\sigma\left(J_i^j\right)$,
as a function of distance $l=|i-j|$ for different RG times $\Gamma$. The 
simulations were run for system size $N=45$ and averaged over $100$ 
realizations. The initial distribution of the bonds is Gaussian with standard 
deviation, $\sigma\left(J_i^j\right)_{\Gamma=0}=\frac{1}{2|i-j|}$ (red straight 
lines in log-log scale). The fields $h_i$ are chosen to be uniformly distributed 
between $0$ and $1$. For initial distributions with exponents $\alpha<0.5$,
the exponent changes and flows to $\alpha=0$ as $\Gamma$ increases.
For exponents, $\alpha>0.5$, the long-distance tails are not altered by the 
flow.   \label{fig:Jdist_Gammadependence} }
\end{figure}

 It is apparent from
the scaling of the source terms that the $l$-dependence of the variance
of the hopping distribution gets modified
throughout the flow, since $l^{-2\alpha}\rightarrow l^{1-4\alpha}$ if 
$\alpha<1/2$. The point $\alpha^{*}=1/2$ is a scaling
fixed point, which is also confirmed
by the sub-leading logarithmic dependence of the variance of the source
terms, $\sigma_{\Delta}^{2}$. Considering parameters slightly away
from this fixed point, $\alpha=\alpha^{*}-\epsilon$, the exponent
generated by the source term is such that $\alpha^{*}-2\epsilon<\alpha$.
Qualitatively, this means that as the RG time scale $\Gamma$ increases,
the source term generates distributions with smaller exponents, which
become the dominant contribution at long distances. Eventually, the
distribution must flow to $\alpha=0$, since $\alpha<0$ is physically
not allowed. In the regime $\alpha>1/2$, on the other hand, we see that the 
source terms
have a distribution with a variance that scales as 
$\bar{\sigma}_{\Delta}^{2}\sim l^{-2\alpha}$.
This means that the source terms do not modify the long distance 
($l\rightarrow\infty$) behavior
of the distribution of $J\left(l\right)$ variables.

In order to check the above argument, we numerically solve
Eqs.~(\ref{eq:J_evol}) and (\ref{eq:h_evol}). The simulations are
done for chains with $N=45$ sites, and the $\Gamma$ parameter flows
from $\Gamma=0$ until $\Gamma=\Gamma_{\mbox{max}},$ where $\Gamma_{\mbox{max}}$
is chosen according to the disorder strength of the hoppings in such
a way that at $\Gamma_{\mbox{max}}$ the energies converge to a fixed value, up
to machine precision. We follow the evolution of both $J_{i}^{j}$ and $h_{i}$ as 
function of $\Gamma$, for chains of $N=45$ sites, and average the results over 
100 disorder realizations.

The standard deviation of the distribution $P_{l,\Gamma}\left(J\right)$, 
$\sigma_{J,\Gamma}\left(l\right)$,
as a function of $l$ for several RG times $\Gamma$ is shown for distinct 
exponents
in Fig.~\ref{fig:Jdist_Gammadependence}. Figures (a) and (b) of 
Fig.~\ref{fig:Jdist_Gammadependence} illustrate that distributions with 
exponents $\alpha<\frac{1}{2}$ flow to  distributions with a constant standard 
deviation, that is, $\sigma_{J,\Gamma}\left(l\right)\sim\text{const}$,
which corresponds to the behavior of $\alpha=0$. At $\alpha=\frac{1}{2}$,
the subleading $\log l$ contribution cannot be seen due to the limitations of 
the system size. In contrast, for $\alpha=0.7$, the long distance power-law behavior of the standard deviation is unaffected by the RG flow, $\sigma_{J,\Gamma}(l) \sim l^{\-\alpha}$, as shown in  Fig.~\ref{fig:Jdist_Gammadependence}(d), in agreement with the previous scaling analysis.

\subsection{Operator Flow \label{sec:Operator-Flow}}

Localization of single particle wave functions can be probed by studying
the flow of single particle operators. One case of particular interest
is the number operator, $c_{i}^{\dagger}c_{i}$, that measures the diffusive 
character of particles in the chain.
We show next that it is possible to study the localized or extended nature
of the system studying the evolution of such operators. 

%We obtain the
%local number operator at the initial RG-time, $\Gamma=0$, is obtained
%in terms of the final operators, $\Gamma=\infty$. 

%$\delta/\left\langle \delta\right\rangle \tilde{\delta}$
%$P\left(\tilde{\delta}\right)$

\selectlanguage{american}%
\begin{figure}
\selectlanguage{english}%
\includegraphics[width=1\linewidth]{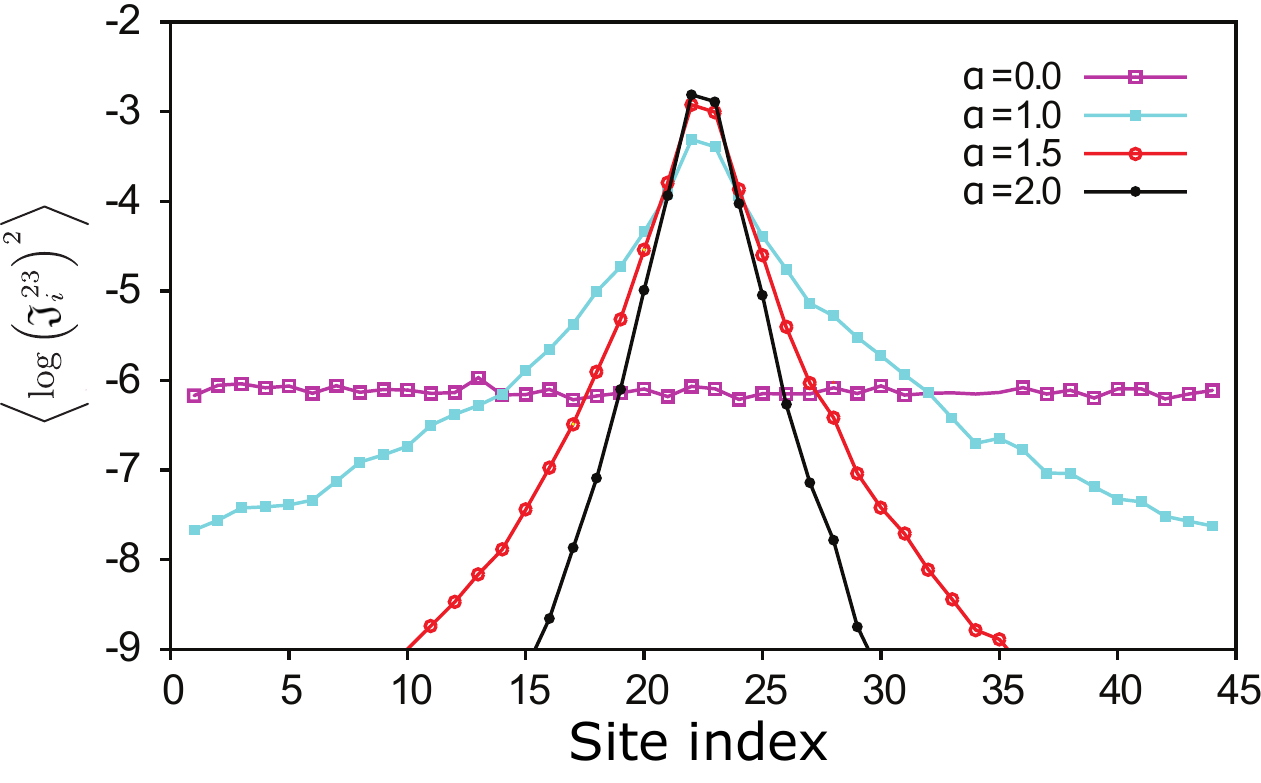}

\selectlanguage{american}%
\caption{\selectlanguage{english}%
Final evolution $\left(\Gamma\rightarrow\infty\right)$ of the number
operator initialized  in the middle of a $45$-site
chain, at site number $23$, $\mathfrak{n}_{23}(\Gamma=0)=c^\dagger_{23}c^{}_{23}$, 
for some representative exponents. At $\Gamma=0$, all $\mathfrak{J}_{23}^{i}$ are 
zero, and only $\mathfrak{h}_{23}$
is equal to one.The asymptotic values are obtained by measuring the final values 
of $\left(\mathfrak{J}_{23}^{i}\right)^{2}$
(see Eq.~(\ref{eq:J_evol-op})). The tilde indicates the set of variables
related to decomposition of the operator flow in terms of an instantaneous basis 
(Eq.~(\ref{eq:op_decomp})).The results are averaged
over 20 disorder realizations. 
\label{fig:operator-evol}\selectlanguage{american}%
}
\end{figure}

\selectlanguage{english}%
As the generator $\eta$ evolves with  $\Gamma$ according to
Eq.~(\ref{eq:flow-generator}), any arbitrary operator in the Hilbert
space also flows, governed by a Heisenberg equation that is analogous to Eq. 
(\ref{eq:HamiltonianRGTimeevolution}).
Let us now consider the evolution of the number operator at site $k$ as a function of the RG time. Writing the local density operator as, $\mathfrak{n}_{k}\left(\Gamma\right)$,
the decomposition in terms of the instantaneous states of 
$n_{k}\left(\Gamma\right)=c_{k}^{\dagger}\left(\Gamma\right)c_{k}
\left(\Gamma\right)$
is 
\begin{equation}
\mathfrak{n}_{k}\left(\Gamma\right)=\sum_{i}\mathfrak{h}_{i}\left(\Gamma\right)n_{i}
+\sum_{\left\langle i,j\right\rangle 
}\mathfrak{J}_{i}^{j}\left(\Gamma\right)c_{i}^{\dagger}c_{j},\label{eq:op_decomp}
\end{equation}
with the initial condition, 
$\mathfrak{h}_{i}\left(\Gamma=0\right)=\delta_{ik}$ and 
$\mathfrak{J}_{i}^{j}\left(\Gamma=0\right)=0$.
We find the general flow equations for
these operator variables to be
\begin{eqnarray}
\frac{d\mathfrak{J}_{i}^{j}}{d\Gamma} & = & 
-J_{i}^{j}x_{j}^{i}\mathfrak{x}_{j}^{i}-\sum_{k=1}^{N}\mathfrak{J}_{i}^{k}J_{k}^{j}x_{
k}^{j}+\sum_{k=1}^{N}J_{i}^{k}\mathfrak{J}_{k}^{j}\left(x_{i}^{k}\right),\label{eq:J_evol-op}\\
\frac{d\mathfrak{h}_{i}}{d\Gamma} & = & 
-2\sum_{k=1}^{N}J_{k}^{i}\mathfrak{J}_{k}^{i}x_{i}^{k},\label{eq:h_evol-op}
\end{eqnarray}
where $\mathfrak{x}_{i}^{j}=\mathfrak{h}_{j}-\mathfrak{h}_{i}$. As 
$\Gamma\rightarrow\infty$, we obtain
$\tilde{n}_{k}$ expressed in the basis of the eigenfunctions of the Hamiltonian.
Since the evolution of the operator variables is intrinsically constraint to the 
couplings of the Hamiltonian,
their flow correlates with the flow of the set $\left\{ h_{i},J_{i}^{j}\right\} 
$.

The flow equations, Eqs.~(\ref{eq:J_evol-op}) and (\ref{eq:h_evol-op})
can be solved numerically. We choose the initial point $k$
to be the midpoint of the chain ($N=22$), and plot the value of 
$\left(\mathfrak{J}_{i}^{k}\right)^{2}$
as a function of the distance $\left|i-k\right|$, averaged over 20 disorder 
realizations. The results are shown in Fig.~\ref{fig:operator-evol}
for different exponents $\alpha$. For large exponents,
$\alpha>1$, the decay is exponential (linear in log scale), indicating
that the density operator stays localized or, equivalently, that the initial 
particle fails to diffuse as a consequence of the
localization of the wavefunctions. For small exponents, $\alpha<1$,
the operator reaches a significant value even at sites arbitrarily
far from the middle, indicating the possibility of long-ranged resonances.
The precise transition point cannot be found due to the restriction
of the system size, but the existence of two phases can already be
inferred. The precise critical point is going to be
discussed later, via other numerical and analytic methods. 

%In this section, we introduced the flow equations as a tool to study
%non-interacting disordered systems. 
%We showed that for the power-law
%hopping models, the distributions of the hoppings flow under the transformations.
%There exists an attractive fixed point at $\alpha=0$, corresponding
%to the GOE ensemble and all exponents below $\alpha=1/2$, flows to
%this fixed point. We showed that for $\alpha>1/2$, the distribution
%of the hoppings remain fixed. The localized or extended nature of
%the wavefunctions can be probed by studying the flow of single particle
%operators. 
One of the handicaps of the flow equation technique, is
that it requires the solution of $\mathcal{O}\left(N^{2}\right)$ coupled differential
equations. This is generally time consuming; the advantage over exact 
diagonalization, however, lies with the ability to extract universal features of 
the system directly from the flow. In the next section, we simplify the flow 
equations further, into a set of decoupled equations, solved sequentially. This 
strong-bond RG method, allows us
to solve the full set of equations efficiently (although still at an $\mathcal{O}(N^3)$ 
cost). It works
in the regime, $\alpha>\frac{1}{2}$, where we show that our assumptions
are correct and the errors accumulated are vanishing in the thermodynamic
limit. We use this method to gain further insights into the delocalization 
transition. 

\selectlanguage{english}%

\section{{\normalsize{}Strong-bond RG method \label{sec:Flow-RG}}}

\begin{figure}
\selectlanguage{english}%
\includegraphics[width=1\linewidth]{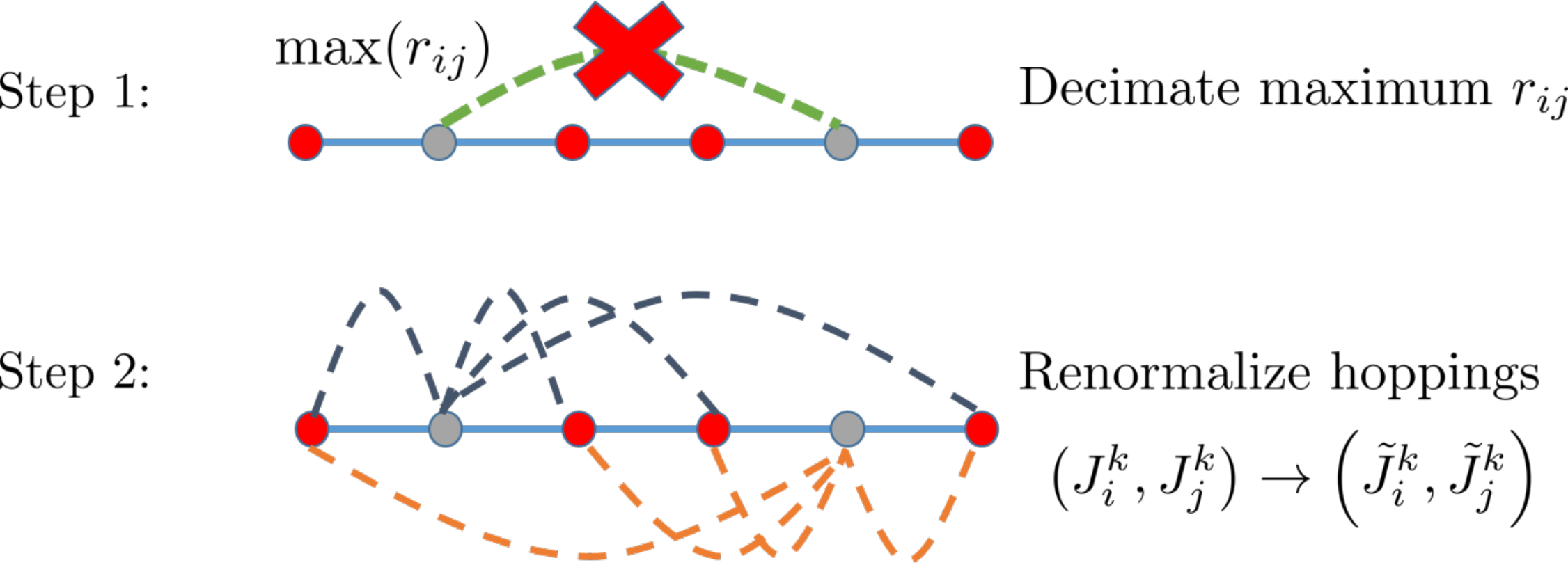}

\selectlanguage{american}%
\caption{\selectlanguage{english}%
(Color Online) Schematic of the steps in the Strong-bond RG method. The first part 
consists
of finding the bond $\left(i,j\right)$ with the maximum $r_{ij}$. Using an 
appropriate unitary, the hopping on the bond is transformed to zero. Hoppings 
connecting to the bond, $\left(\tilde{J}_{i}^{k},\tilde{J}_{j}^{k}\right)$, and 
fields on its sites $\left(\tilde{h}_{i},\tilde{h}_{j}\right)$, are 
renormalized.
This procedure is iterated until all bonds are set to zero. The strong disorder 
allows is to make a crucial simplification: Once a bond is set to zero, we 
neglect its regeneration in subsequent steps. This produces a negligible error 
if the generated $\tilde{r}_{ik}$ and
$\tilde{r}_{jk}$ are smaller than the removed $r_{ij}$. After 
$\mathcal{O}\left(N^{2}\right)$
steps, where $N$ is the system size, the Hamiltonian is diagonal.
\label{fig:Schematic-RG}\selectlanguage{american}%
}
\end{figure}

The exact two-site solution allows us to devise an RG-scheme of sequential
transformations. These transformations produce an alternative scheme for 
constructing the unitary that diagonalizes the Hamiltonian, and it can also 
efficiently yield 
an approximate solution of the flow in Eqs.~(\ref{eq:J_evol}) and 
(\ref{eq:h_evol}). As we
noted in Section \ref{sub:Two-site-sol}, the FE diagonalizes the
two-site problem with a characteristic RG timescale, 
$\tau_{\Gamma}\sim\frac{1}{r^{2}}$.
This suggests an approximate solution to the $N$-site problem by
breaking it into a sequence of two-site rotations ordered by the
magnitude of $r$. Each rotation sets the hopping across the bond to zero. At 
every RG step, we transform the bond given with
the largest value of $r$ and renormalize the bonds connected to sites
of the decimated bond. In Fig.~\ref{fig:Schematic-RG}, we schematically
show the RG procedure.

This RG procedure can be interpreted as an ordered sequence of two-site
rotations, analogous to the Jacobi algorithm used to diagonalize matrices  
\cite{GolubLoanbook}.
The difference from the Jacobi rotation method is that the FE provides
a natural ordering the decimations, the descending value of $r$. 

\selectlanguage{american}%

\selectlanguage{english}%
The strong-bond RG procedure relies on the two-site transformation, 
Section~\ref{sub:Two-site-sol}. In practice, we employ the 2-site transformation 
as a Jacobi rotation on the entire Hamiltonian. The guidance provided from the 
flow-equations is the order in which we should pursue the transformations. 
Relegating the details of the rotations to 
App.~\ref{sec:Appendix-Flow-RG:-Derivation}, we provide here the resulting RG 
procedure steps:
\begin{enumerate}
\item Find the largest non-decimated $r$, say $r_\textrm{max}=r_{ij}=\sqrt{4\left(J_{i}^{j}\right)^2+x_{ij}^2}$, between 
sites
$\left(i,j\right)$. 
\item Compute the corresponding bond angle
\begin{equation}
\alpha_{ij}=\mbox{sgn}\left(J_{i}^{j}x_{i}^{j}\right)\frac{\theta_{i}^j}{2},
\label{eq:maxrbondangle}
\end{equation}
where
\begin{equation}
x_{i}^{j}=h_{j}-h_{i},
\label{eq:xdefinition}
\end{equation}
and
\begin{equation}
\theta_{i}^j=\arctan\left|\frac{2J_{i}^{j}}{x_{i}^{j}}\right|.
\label{eq:thetadefinition}
\end{equation}

\item Set the corresponding $J_{i}^{j}$ to zero.
\item Renormalize all bonds connected to sites $i$ or $j$ according to:
\begin{eqnarray}
\tilde{J}_{i}^{k} & = & 
J_{i}^{k}\cos\left(\alpha_{ij}\right)+J_{k}^{j}\sin\left(\alpha_{ij}\right),
\label{eq:FlowRG-Jrenorm-1}\\
\tilde{J}_{j}^{k} & = & 
-J_{i}^{k}\sin\left(\alpha_{ij}\right)+J_{j}^{k}\cos\left(\alpha_{ij},
\right).\label{eq:FlowRG-Jrenorm-2}
\end{eqnarray}
where $\alpha_{ij}$ was defined in Eq. (\ref{eq:maxrbondangle}).
\item Renormalize the fields $h_{i}$ and $h_{j}$ according to
\begin{eqnarray}
\tilde{h}_{i,j} & = & \frac{1}{2}\left[H_{ij}\pm 
r_\textrm{max}\mbox{sgn}\left(x_{i}^{j}\right)\right],\label{eq:FlowRG-hrenorm}
\end{eqnarray}
 where $H_{ij}=h_{i}+h_{j}$.
\item Compute the renormalized values of $r$: $\tilde{r}_{ik}$ and 
$\tilde{r}_{jk}$.
\end{enumerate}

The number of steps until the Hamiltonian becomes diagonal scales
as $\mathcal{O}(N^2)$, where $N$ is the system size. Each step renormalizes $\mathcal{O}(N)$ bonds connected to the decimated bond. Therefore, the number of computations necessary to compute all eigenvalues using this method is $\mathcal{O}(N^3)$. Also, in this RG proposal,
each diagonal element, that converges to the approximate eigenvalue,
is renormalized $\mathcal{O}\left(N\right)$ times. This is an advantage
in comparison to other proposals, like the one by Javan Mard \textit{et
al.\cite{JavanmardPRB}, }for example, if one is interested in level
spacing. In the latter RG proposal, sites, and not bonds, are removed
from the chain. This procedure also coincides with the procedure in 
Ref.~\onlinecite{RademakerPRL}, which was developed simultaneously, and applied 
to many-body systems. 

\subsection{Universal properties from the strong-bond RG}

The strong bond renormalization group approach primarily provides a new 
perspective from which the universal properties of disordered quantum systems 
could be extracted.
First, the successive RG transformations suggest representing the problem as a 
2-dimensional scatter plot on the $x-J$ plane. Each point in the plot 
corresponds to a particular bond connecting two sites, say $i$ and $j$. Its ``y'' 
value is the  bond strength
$J_{i}^{j}$, and its ``x'' value is the difference of the on-site fields $\left\{ 
h_{i},h_{j}\right\} $,
$x_{i}^{j}=h_{j}-h_{i}$. A diagonal Hamiltonian, for example,
would correspond to having all points on the $x_{i}^{j}$ axis. 

\begin{figure}
\includegraphics[width=1\linewidth]{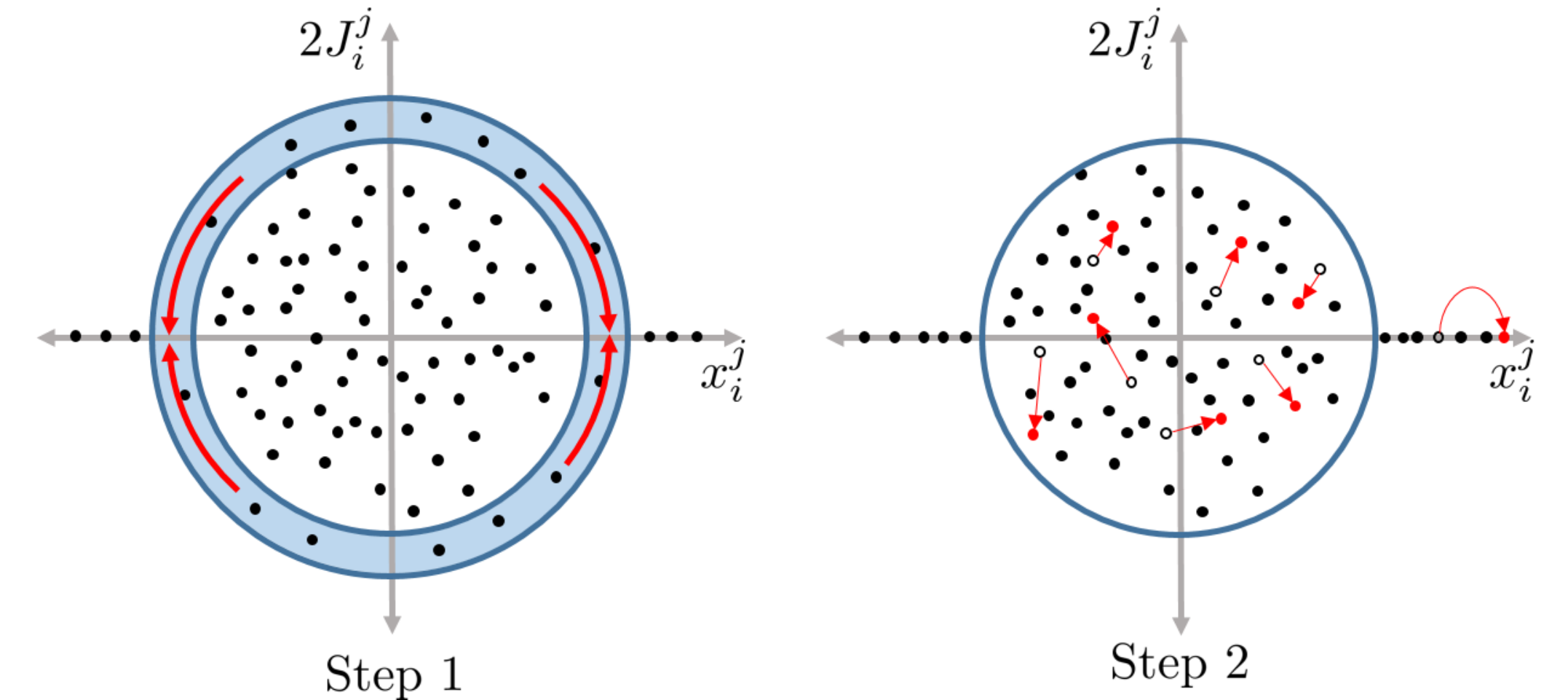}
\caption{(Color Online) The representation strong bond RG procedure in the $x-J$ 
space. Each point represents a bond, and its distance from the origin is 
$r_{ij}$. the strong bond RG rotates the bonds within a large-$r$ shell. In the 
first
step, the bonds with largest $r_{i}^{j}$ are rotated to the $x_{i}^{j}$
axis. Next, the bonds connected to the decimated bond
undergo a renormalization via \textbf{ }Eqs.~(\ref{eq:FlowRG-Jrenorm-1}), 
(\ref{eq:FlowRG-Jrenorm-2}),
and (\ref{eq:FlowRG-hrenorm}). We perform one approximation: once eliminated, a 
bond
is not allowed to assume finite values again, and these points which lie on the 
$x$-axis beyond the $r$-cutoff move horizontally only.\textbf{ 
\label{fig:RG-scheme-xJ}}}
\end{figure}

The emerging picture provides a convenient way to represent the RG flow of the 
coupling distribution under the scheme of the previous Section, 
\ref{sec:Flow-RG}.
As shown schematically in Fig.~\ref{fig:RG-scheme-xJ},
a decimation corresponds to rotating bonds in the largest circular
shell, bringing them to the $x_{i}^{j}$ axis. In the later
steps, the points within
the circle get modified according to the Eqs.~(\ref{eq:FlowRG-Jrenorm-1}),
(\ref{eq:FlowRG-Jrenorm-2}), and (\ref{eq:FlowRG-hrenorm}). Let
us call it $P_{\Gamma}\left(x\right)$ the distribution of $x_{i}^{j}$
at scale $\Gamma$. As one decimates all the bonds in the Hamiltonian,
the final distribution of points on the $x_{i}^{j}$ axis is obtained.
The final distribution, $P_{\Gamma\rightarrow\infty}\left(x\right)$,
is the distribution of the level separations for all the eigenvalues
of the Hamiltonian. It is proportional to the level correlation function 
\cite{mehtaRMT}
which, in the limit $x\rightarrow0$, is identical with the level
spacing statistics. For simple localized and extended states it is given by
\begin{equation}
\lim_{x\rightarrow\infty}P_{\Gamma\rightarrow\infty}\left(x\right)\propto\begin{cases}
\text{ const.}, & \text{if the phase is localized}\\
x, & \text{if the phase is extended}.
\end{cases}
\end{equation}

Examining the long RG-time fixed points
of the flow of the distributions, therefore, allows us to identify the different 
phases of a system, and extract their universal properties.

The $x-J$ space gives an intuitive picture for how the level-spacing 
distributions emerge in the two fixed points of the PBRM model - the localized 
and extended phases. 
A level repulsion, as in the extended phase, is obtained from a uniform 
distribution of bonds in the $x-J$ space
of Fig.~\ref{fig:RG-scheme-xJ}. In contrast, for localized
states that do not repel each other, the joint distribution has a
finite range in the phase space with a length scale $\xi\ll r_\textrm{max}$.
As a simplification, we assume that the effect of the bond renormalization,
which is schematically represented in Fig.~\ref{fig:RG-scheme-xJ},
can be ignored. First consider the case of a uniformly distributed
bonds in the phase space, $P_{\Gamma}\left(J,x\right)\sim\text{const.}$ In
this case, the number of decimations in a circular shell of radius
$r_{\text{max}}$ and width $dr_{\mbox{max}}$ is $N_{\text{dec}}\propto2\pi 
r_{\text{max}}dr_{\text{max}}$.
The number of decimations fixes the distribution of bonds at  
$x=r_{\text{max}}$.
Therefore, we have the distribution

\begin{equation}
P_{\Gamma}\left(x\right)dx\sim r_{\text{max}}dr_{\text{max}}\sim 
xdx,\label{eq:P(x)delocalized}
\end{equation}
which correctly reproduces the Wigner-Dyson statistics in the limit
of small level spacing. Now, we can repeat the same analysis for 
$P_{\Gamma}\left(J,x\right)\sim e^{-J/\xi}$.
In this case, the number of decimations goes as $N_{\text{dec}}\propto\xi 
dr_{\text{max}}$.
Consequently, we have for the distribution of level separations

\begin{equation}
P_{\Gamma}\left(x=r_{\text{max}}\right)\sim\text{const},
\end{equation}
consistent with Poisson statistics for localized states at the small
level spacing limit. We note that this analysis relies on the assumption
that the renormalization of the bonds does not significantly alter the marginal
distribution of $J$. In the following, we show that this approximation
is reasonable. Note that the bond distribution function in the $x-J$ space 
typically separates into a product distribution, with a uniform distribution on 
the x-axis at late stages of the flow. The J-distribution then arbitrates the 
level statistics: If it is uniform, we obtain Wigner-Dyson statistics, and if it 
is concentrated near $J=0$, a Poisson-type distribution emerges.

\subsection{Strong-bond RG and the delocalization transition\label{RGres}}

Let us consider the effects of bond renormalization on the
marginal bond distribution, $P_{\Gamma}\left(J\right)$ of the PBRM model. 
Examining Eqs.~(\ref{eq:FlowRG-Jrenorm-1})~and~(\ref{eq:FlowRG-Jrenorm-2}),
the evolution of the bonds $J$ may be interpreted as a random walk
with an amplitude proportional to $J$. To be more precise, the variance
of the bonds change under renormalization as
\begin{equation}
\sigma^{2}\left(\tilde{J}_{i}^{k}\right)\approx\left\langle 
\left(J_{i}^{k}\right)^{2}\right\rangle +\left\langle 
\left(J_{j}^{k}\right)^{2}-\left(J_{i}^{k}\right)^{2}\right\rangle 
\sin^{2}\alpha_{ij}\end{equation}
where we have assumed that the product $J_{i}^{k}J_{j}^{k}$ is uncorrelated,
$\left\langle J_{i}^{k}J_{j}^{k}\right\rangle \sim\left\langle 
J_{i}^{k}\right\rangle \left\langle J_{j}^{k}\right\rangle =0$. The rotation angle $\alpha_{ij}$ is defined in Eq. (\ref{eq:maxrbondangle}).
The change of the standard deviation is reminiscent of a one-dimensional
random walk with a variable amplitude for each of the steps. Furthermore, we can 
assume that the two bonds that are renormalized, $J_i^k,\,J_j^k$ are of 
comparable range. The change in variance is 
then:
\begin{equation}
\left|\Delta\left\langle \left(J_{i}^{k}\right)^{2}\right\rangle 
\right|\sim\left\langle 
\left|\left(J_{i}^{k}\right)^{2}-\left(J_{j}^{k}\right)^{2}\right|\right\rangle 
\sim\left\langle \left(J_{i}^{k}\right)^{2}\right\rangle ,
\end{equation}
where, note that the average change is non-zero because we are computing the magnitude.
So the random change in the magnitude of the bond is proportional to the bond 
strength itself. Relying on this insight, we can model the flow of the $J$ 
distribution as a diffusion equation with a $J$-dependent diffusion constant, 
$D\left(J\right)=D_{0}J^{2}$. Before writing the equation, we note that the sum of undecimated couplings,
$\sum\limits_{i\neq j}\left(J_i^j\right)^2$, remains constant throughout the RG flow. We 
account for that by adding a rescaling term in the diffusion equation. The 
combined equation is then:
\begin{equation}
\frac{\partial 
P_{\Gamma}\left(J\right)}{\partial\Gamma}=\frac{\partial}{\partial 
J}\left(D_{0}J^{2}\frac{\partial P_{\Gamma}\left(J\right)}{\partial J}-\gamma J 
P_{\Gamma}\left(J\right)\right),\label{eq:diffusionansatz}
\end{equation}
where the values of the diffusion constant $D_{0}$ depend on the details of the 
distributions of $J$ and $x$ at the renormalized scale. $\gamma$ is a Lagrange 
multiplier which is adjusted to maintain the variance of the problem constant. 

The steady states of Eq. (\ref{eq:diffusionansatz}) are easy to infer. From the 
structure of the diffusion equation we see that the solutions must be scale 
invariant, i.e., power-law distributions. For any power low distribution, 
\begin{equation}
P_{\Gamma}\left(J\right)\sim C\left(\Gamma\right)J^{-\beta},
\end{equation}
the exponent $\beta$ would remain invariant. Furthermore, since $\gamma$ is 
adjusted to maintain the variance of $J$ constant, the $\gamma$ rescaling term 
would actually make any power-law distribution a fixed point.

The discussion above makes us consider what appears to be the most crucial 
feature of the PRBM. The initial hopping distribution $P_{\Gamma=0}(J)$ for the 
power-law decaying random hopping is already a power law for almost all $J$'s. 
Therefore, it is a fixed-point distribution from the start. In more detail, the 
initial marginal bond distribution of all bonds $P_{\Gamma=0}\left(J\right)$ for 
a length $N$ chain has two distinct behaviors. 
At small $J$'s, with $J<J_c= \frac{1}{N^{\alpha}}$  it is uniform, and for 
$J>J_c$, it is a power law:
\begin{equation}
P_{\Gamma=0}\left(J\right)\propto\begin{cases}
\frac{1}{J^{1+1/\alpha}} & ,\text{for }J>J_{c}\\
N^{\alpha} & ,\text{for }J<J_{c}
\end{cases}
\label{eq:P(J)atGamma=0}
\end{equation}
This is calculated and numerically verified in 
Appendix~\ref{sec:Initial_hoppings}.
Since each of the two segments is a power-law, both remain invariant. The 
crossover range, however, may change in the flow. Any changes of $J_c$ during 
the flow, however, are bound to result in a scale-invariant change. Therefore, 
we assume that $J_c\sim 1/N^{\alpha}$ throughout the flow. This expectation is 
confirmed by our numerics, as discussed in the next section.

\begin{figure*}
\includegraphics[width=1\linewidth]{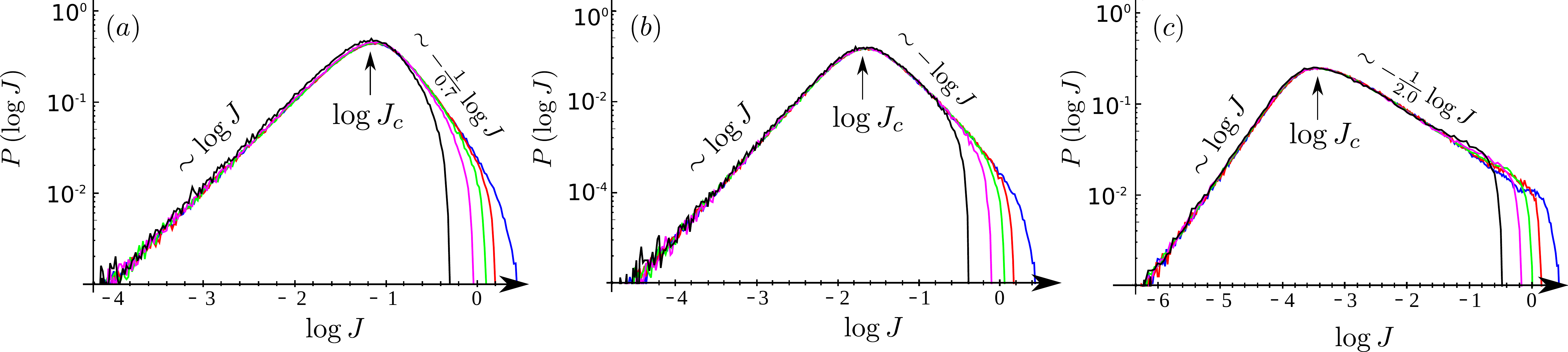}
\caption{(Color Online) Marginal distribution $ P_\Gamma \left(\log J\right)$ in 
the log scale for different RG steps $\Gamma$.  From (a) to (c), we plot the 
evolution of the marginal distribution for exponents $\alpha=0.7$, $\alpha=1.0$ 
and $\alpha=2.0$. Different colors represents different RG steps; $\Gamma=1$, 
$1000$, $2000$, $3000$ are represented by blue, red, green, magenta, and black, 
respectively. As seen from Eq.~(\ref{eq:P(J)atGamma=0}) there are two distinct 
regimes in the probability distribution. The crossover scale is given by $J_c$. 
Below this, $J<J_c$, $P_\Gamma \left(\log J\right)\sim \log J$ and above the 
scale, for $J>J_c$, $P_\Gamma \left(\log J\right)\sim -\frac{1}{\alpha}\log J$. 
We note that as the bonds are decimated, the behavior of the distribution below 
and above the crossover remains unchanged. The system has size $N=100$ and we 
average over 20 disorder realizations. \label{fig:P-J-RG-flow}}
\end{figure*}

Now we can address the critical behavior of the power-law hopping problem. The 
fact that any power-law marginal $J$ distribution is also marginal in the RG 
sense implies that the entire $\alpha>0.5$ parameter range is critical. The 
transition, we show, emanates from the size dependence of the marginal 
distribution. 
The two regimes of $P_{\Gamma}(J)$ also imply two regimes of level spacings. 
After very little flow, the marginal $x$ distribution flattens, and the full x-J 
bond density is:
\begin{equation}
P(x,J)\approx c \frac{r_\textrm{max}}{r_0}N^2 P_{\Gamma}(J)
\end{equation}
where $c$ is a constant, and $r_\textrm{max}$ is the RG cutoff, and $r_0$ is the largest RG cutoff. As we transform 
away bond in the arc $r_\textrm{max}-dr<r<r_\textrm{max}$, and reduce 
$r_\textrm{max}$, the number of bonds affected, and hence the density of level 
spacings is:
\begin{equation}
\rho(x)\sim\left\{\begin{array}{cc} 
\textrm{const}, & r>J_c \\
c\,\,r & r<J_c
\end{array}
\right. \label{dens-eq}
\end{equation}

\textbf{}%

Next, we need to find out how the mean level spacing, $\bar{\delta}$, scales. 
For $\alpha>1/2$, we expect $\bar{\delta}_{N}\sim\frac{1}{N}$, since the 
system's bandwidth is size independent. In Appendix~\ref{sec:app-bandwidth} we 
demonstrate this result under the flow-equations scope. Alternatively, we can 
use the fact that the bandwidth of the Hamiltonian, $W$, is bounded by the norm 
of the off-diagonal terms, added to the disorder width $w_0$ : 
\begin{equation}
W\leq w_0+\sqrt{\int_{0}^{N}dl\ J_\textrm{typ}^{2}\left(l\right)}\propto 
N^{\frac{1}{2}-\alpha}+\text{const.},
\end{equation}
where $J_\textrm{typ}\left(l\right)\propto\frac{1}{l^{\alpha}}$, are the length-dependent
hopping terms. In the thermodynamic limit, when $\alpha>1/2$, the 
length-dependent correction vanishes.

The phase of the system, and the delocalization transition, are inferred from the 
level-spacing statistics, expressed in terms of the rescaled level spacing. We 
denote the rescaled level spacing as $\bar{\epsilon}=\epsilon/\bar{\delta}_N$. As Eq. (\ref{dens-eq}) shows, 
level repulsion appears below the energy difference $J_c$. In terms of the 
scaled level spacing, this implies that level-repulsion sets in for rescaled 
energy difference:
\begin{equation}
\bar{\epsilon}_c(N)\approx \frac{{J_c}}{{\bar{\delta}_{N}}}\sim N^{1-\alpha}.
\end{equation}
For $\alpha>1$, $\bar{\epsilon}_c$ vanishes in the thermodynamic limit. When 
$\alpha\leq 1$, the crossover point $J_{c}$
is non-negligible in the thermodynamic limit. When the decimation
scale reaches $J_c\sim 1/N^{\alpha}$, the distribution of bonds becomes
uniform in the $J-x$ phase space. For $\alpha<1$, the level repulsion emerges at  $\epsilon_c\sim N^{1-\alpha}\bar{\delta}_{N}$, which is  much larger than the average level spacing. On the other hand, for $\alpha>1$, $\epsilon_c\ll \bar{\delta}_{N}$ and, therefore, the distribution of level spacings is Poissonian. The phase diagram of 
Fig.~\ref{fig:Phase-diagram} emerges naturally from the strong-bond RG 
analysis. 

The strong-bond RG picture also yields the correlation length scaling of the 
transition. Let us define $\xi$ as the chain length that allows us to determine 
the phase of the system from the level-statistics distribution. In the 
delocalized phase, $\alpha<1$, we would require $\epsilon_c(\xi)> a 
\bar{\delta}_{\xi}$, with $a>1$ being some constant, which we could set to be 
$a=2$ without loss of generality. This would imply $\xi^{1-\alpha}=a$, and:
\begin{equation}
\xi_{del}\sim a^{1/(1-\alpha)}.\label{xi1}
\end{equation}
Similarly, in the localized phase, $\alpha>1$, level repulsion will always 
emerge at some finite energy scale, as the scaling of $\epsilon_c(N)$ suggests. 
This scale, however, must be well below the average level spacing. We would then 
require $\epsilon_c(\xi)< \bar{\delta}_{\xi}/a$. This leads to:
\begin{equation}
\xi_{loc}\sim a^{1/(\alpha-1)}.\label{xi2}
\end{equation}
Together, Eqs. (\ref{xi1}) and (\ref{xi2}) imply:
\begin{equation}
\ln\xi\sim \frac{1}{|\alpha-1|}
\end{equation}
which is consistent with the results of Ref.~\onlinecite{MirlinPRE}. In this analysis, we note that the localization length for $\alpha>1$, becomes the correlation length in the delocalized regime, for $\alpha<1$.

\subsection{Numerical results}

\begin{figure}
\includegraphics[width=1\columnwidth]{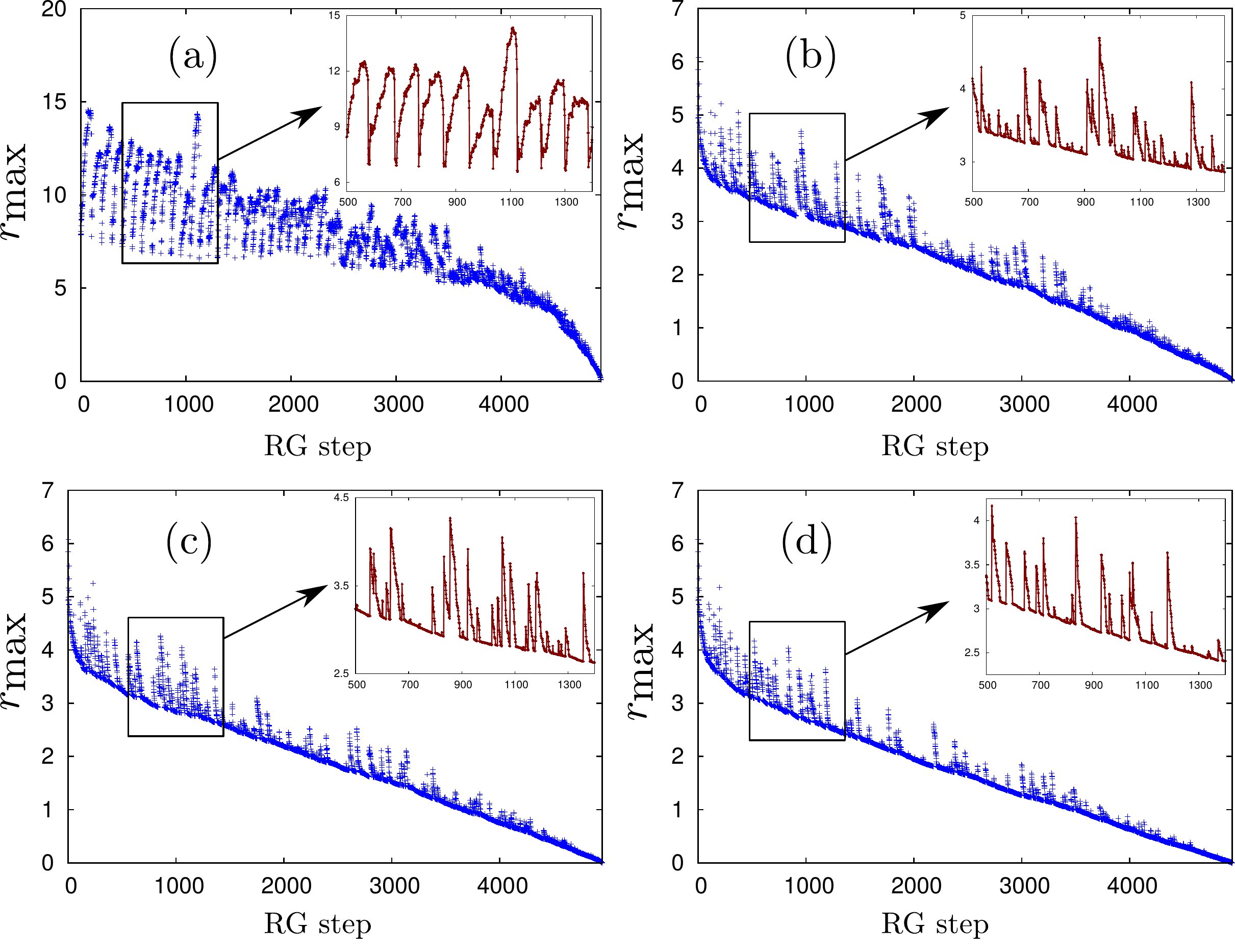}

\caption{Decimated $r=r_{\mbox{max}}$ as function of the RG step in a given
disorder realization, for distinct exponents, (a)~$\alpha=0.1$, 
(b)~$\alpha=0.7$, (c)~$\alpha=1.0$, and (d)~$\alpha=2.0$. The behavior of the 
slopes of the peaks in
the curves differs significantly, as in (a) $r$ increases in several
consecutive RG steps, while in (b-d) a bond that is generated with 
$r>r_{\text{max}}$
 is immediately removed. Notice additionally
that, in contrast with (a), in (b-d) the decimated $r_{\text{max}}$ decreases 
consistently during the RG flow.
\label{fig:Decimated-r-typ}}
\end{figure}

The scaling statements made above in Sec. \ref{RGres} were confirmed 
numerically.
In Fig.~\ref{fig:P-J-RG-flow}, we plot the marginal distribution
$P\left(J\right)$ for different RG steps. Clearly, when 
$r_{\text{max}}>J_{\text{cross}}$,
the exponent of the initial power law remains unchanged. In contrast, below the 
cutoff
the bonds are uniformly distributed.

The method is reliable for $\alpha>0.5$, it is asymptotically accurate as 
$\alpha\rightarrow\infty$
when all states are localized, and it fails in the strongly delocalized regime, 
$\alpha<0.5$. 
The failure in the $\alpha<0.5$ region can be traced to the approximation that a 
transformed bond is not regenerated: once removed, the
corrections to a transformed bond are neglected in later RG steps. This 
assumption is crucial for the formulation of an RG flow, since such flow relies on 
a decreasing scale, $r$. This approximation, however, breaks down when one of 
the renormalized
bonds, $\tilde{r}_{ik}$ or $\tilde{r}_{jk}$, is greater than $r_{ij}$.
Such ``bad decimations'' correspond to cases when
delocalized clusters of three or more sites should be diagonalized
simultaneously.  

The numerical implementation of our RG method can also be used to obtain the 
eigenvalues of particular realizations of the problem. 
Fig.~\ref{fig:Decimated-r-typ} contrasts the evolution of $r_\textrm{max}$ 
during the RG flow for the different phases. For
 $\alpha<\frac{1}{2}$, the $r$ values of transformed bonds increase as a 
function of the decimation step (Fig.
\ref{fig:Decimated-r-typ}a). Indeed, in this regime the two-site solution is not applicable; the full flow equations of Section \ref{sec:Method:-Wegners-Flow}, that can describe macroscopically large clusters, are needed. This effect, however, is absent for  $\alpha>0.5$, 
including at the
transition point, $\alpha=1$. In those cases, RG steps occasionally generate a 
family
of large $r$'s. But these $r$'s are promptly eliminated, and $r_\textrm{max}$ 
continues to monotonically decrease, and the  method is controlled.

\begin{figure}
\includegraphics[width=1\columnwidth]{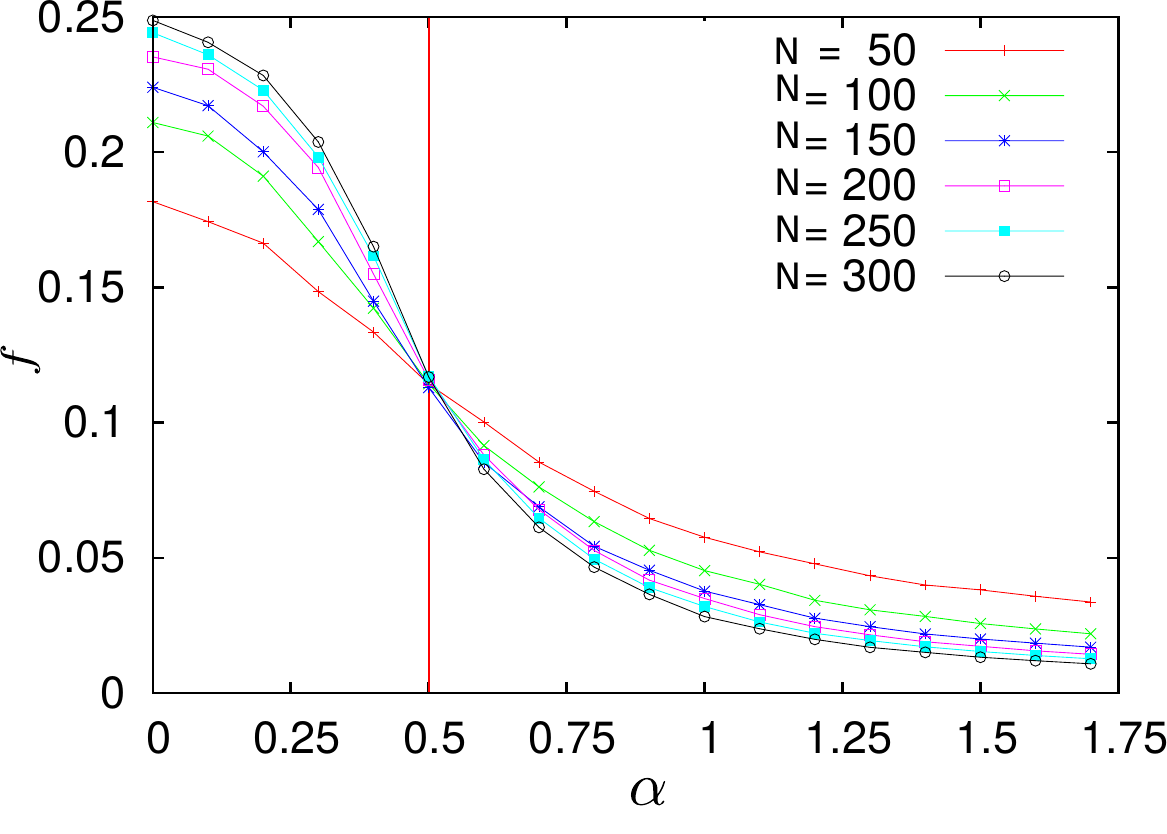}

\caption{Fraction of decimations $\left(\,f\,\right)$ that does not lower
the energy scale $r$, in the Strong-bond RG scheme. There is a transition
at $\alpha=0.5$, indicating the failure of the Strong-bond RG for $\alpha<0.5$.
The Strong-bond RG has vanishing fraction of decimations in the thermodynamic
limit for $\alpha>0.5$. \label{fig:baddecimation-finitesizeeffect}}
\end{figure}

We also considered the number of bad decimations as a function of system size. 
Remarkably,
the fraction of bad decimations vanishes in the thermodynamic
limit for all $\alpha>0.5$, as shown in 
Fig.~\ref{fig:baddecimation-finitesizeeffect}.
Our plot shows a crossing at $\alpha=\frac{1}{2}$, which is the transition
point from intermediate level statistics to GOE level statistics.
This figure reveals that the method fails only for the strongly delocalized
part of the phase diagram. The RG procedure is valid at and around the 
localization-delocalization
critical point, $\alpha=1$ and therefore provides an accurate description
of the critically delocalized wavefunctions. 

\begin{figure}
\includegraphics[width=1\linewidth]{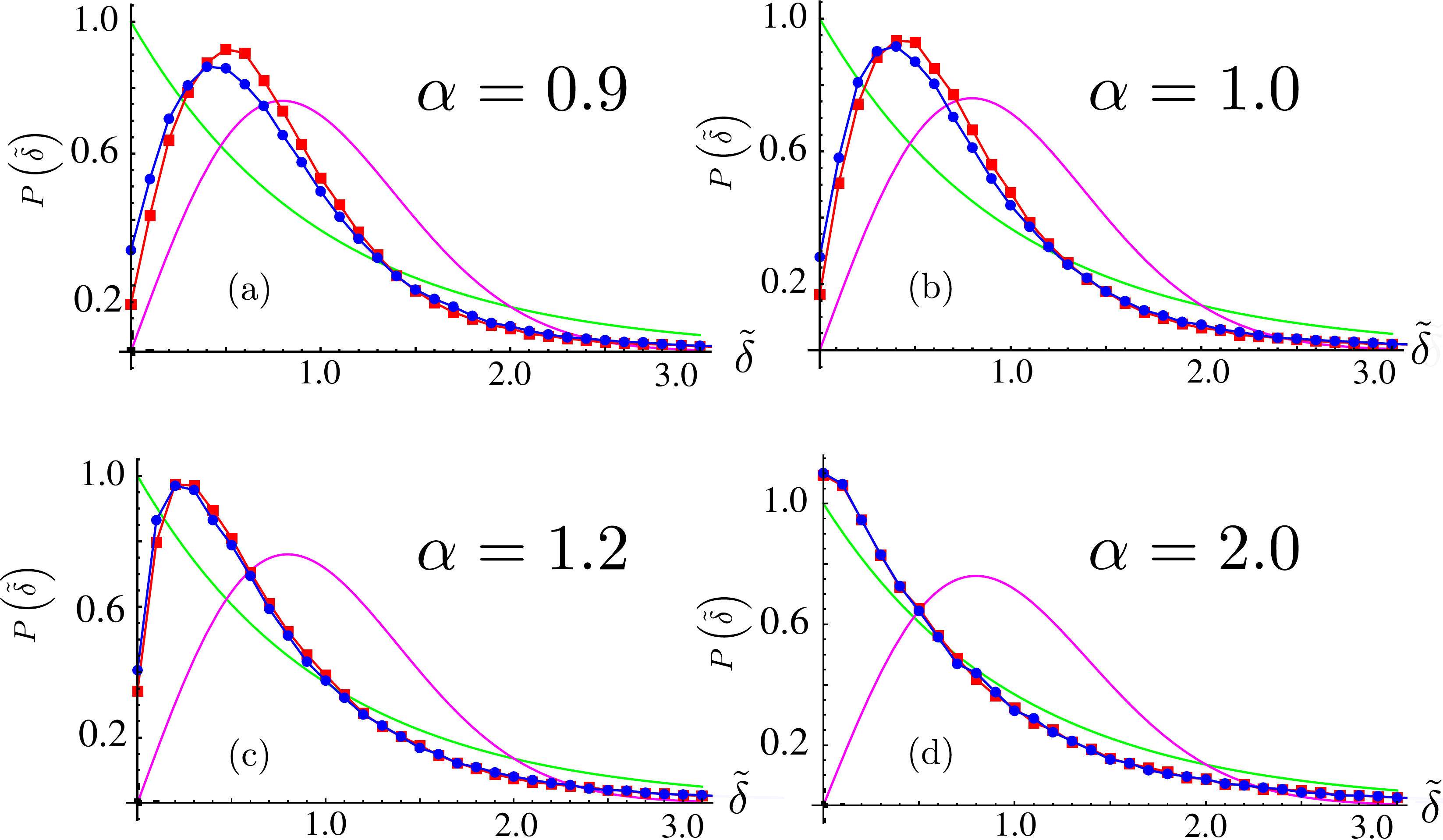}

\caption{(Color Online) Level spacing comparison for eigenvalues obtained 
through Strong-bond RG (blue
circles) and exact diagonalization (red squares), normalized by the mean 
level spacing value. (a), (b), (c), (d)
correspond to exponents $\alpha=0.9$, $1$, $1.2$, and $2$, respectively. For comparison,
we also plot the analytical expressions for Poisson (green) and Wigner-Dyson
(magenta) statistics. The system size is $N=400$
sites, averaged over $500$ disorder realizations. The $G_{i}^{j}=J_{i}^{j}\left|i-j\right|^{\alpha}$ and $h_{i}$ 
variables follow Gaussian distributions, with unit standard derivation.
 \label{fig:levelspacingstat-FLowRG}}
\end{figure}

In Appendix \ref{sec:Appendix-Flow-RG:-Derivation}
we compare the single particle spectrum and its level statistics obtained from 
exact diagonalization
and the RG procedure in the regime of applicability, $\alpha>0.5$. We see a 
decent agreement between the strong-bond RG results and exact diagonalization 
for a variety of $\alpha$ values. We considered chains of 400
sites, and averaged over 500 disorder realizations. We choose $G_{i}^{j}$ and 
$h_{i}$ to have Gaussian distributions, with unit standard derivation. The level 
spacings,
$\delta$, are computed in units of their mean value. It is well known
\cite{wigner1951,BeenakkerRMP,mehtaRMT,monvel1995} that random matrices
in the GOE ensemble have a universal distribution for the level spacing, 
$P\left(\delta\right)=\frac{\pi\delta}{2}\exp\left(-\frac{\pi}{4}\delta^{2}
\right)$.
In contrast, localized Hamiltonians exhibit no level repulsion and
hence the level-spacing statistics are Poissonian, 
$P\left(\delta\right)=\exp\left(-\delta\right)$.
As discussed in Section~\ref{sec:The-model-PBRM}, the critical point
 at $\alpha=1$ exhibits intermediate level-statistics
that are neither Poisson nor Wigner-Dyson. This feature of the critical 
level-spacing
statistics can be reproduced using the strong-bond RG, as shown in 
Fig.~\ref{fig:levelspacingstat-FLowRG}(b)
for the critical point. In contrast, for $\alpha=5$, the system is
localized at all eigenvalues, and hence the level-spacing statistics
are Poisson as shown in Fig.~\ref{fig:levelspacingstat-FLowRG}(d).
Slightly away from the critical point at $\alpha=0.9$,
Fig.~\ref{fig:levelspacingstat-FLowRG}(a),
we see that there is a deviation for small $\delta$ in the level
repulsion obtained using exact diagonalization and the Strong-bond RG procedure.
We attribute this deviation to finite-size effects, which were anticipated in 
Sec. \ref{RGres}. We find, therefore, strong support for all aspects of the 
strong-bond RG analysis from the numerical results.  

\begin{figure*}
\includegraphics[width=1.\linewidth]{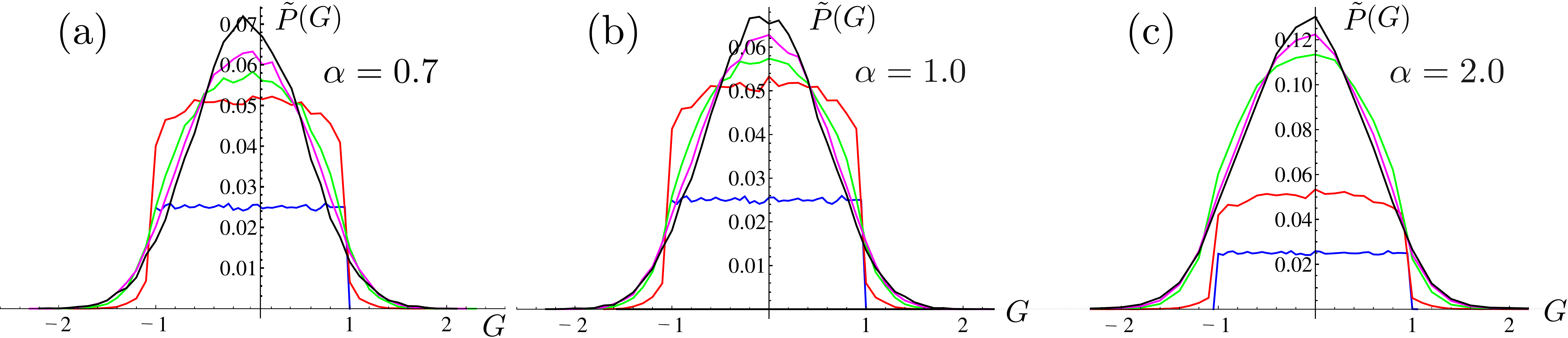}

\caption{(Color Online) Distribution of non-decimated distance-independent bond 
couplings $G=J_{i}^{j}\left|i-j\right|^{\alpha}$,
$\tilde{P}\left(G\right)$, as the RG flows, at RG steps $N_{\textmd{steps}}=1$
(blue), 100 (red), 1000 (green), 2000 (magenta), and 3000 (black).
The number of sites is $N=100$, and the total number of steps
to diagonalize the Hamiltonian is $N_{\textmd{steps}}=4950$. The exponents
shown are (a)$\alpha=0.7$, (b) $\alpha=1.0$, (c) $\alpha=2$. In
all cases, the initial distribution of $G$ is uniform, from $-1$
to $1$ (blue curve). At later RG steps, the $G$ distribution becomes
Gaussian.\label{fig:non-dec-FlowRG}}
\end{figure*}

We also observe some universal behavior of the distributions under
the RG procedure. Since $\alpha$ remains fixed during the RG flow, we can study 
the behavior of the distributions of $G=J_{i}^{j}\left|i-j\right|^{\alpha}$.
In Fig.~\ref{fig:non-dec-FlowRG}, we plot the distribution of these
bonds $P\left(G_{i}^{j}\right)$ as a function of the decimation
step. We illustrate with the case when the initial distributions of the
bonds $G_{i}^{j}$ are uniform (from 0 to 1), but we have verified that the 
behavior is similar if $G_{i}^{j}$ has a Gaussian distribution. Under the RG 
procedure, after a large
number of steps, the bonds become normally distributed. This is a
feature not only at the critical point but also away from
it. We note that Levitov\cite{Levitov2} predicted that the fixed
point distribution of bonds is a normal distribution using a real-space RG 
scheme. We see that the same holds true for the strong-bond
RG procedure.

This method also provides us the eigenfunctions of the Hamiltonian.
Since each decimation corresponds to a rotation of the basis, the
full unitary matrix for diagonalizing the Hamiltonian can be obtained
from the product of all the two-site decimations. The eigenfunctions
obtained using this method have remarkably close behavior to the exact
eigenstates. In the Appendix \ref{sec:Wavefunction-and-IPR}, we outline
the procedure to obtain the full eigenfunctions of the system. We
also calculate the critical, fractal dimensions from the inverse participation ratio (IPR) statistics.
This indicates that the method is quite controlled and gives us the
correct behavior at the critical point.

\section{{\normalsize{}Conclusions}}

In this paper, we have shown that the Wegner's flow equation is a
very useful tool to study  localization transitions . We choose,
for concreteness, the example of non-interacting  particles with power-law
decaying hoppings. This method allows us to map out the
phase diagram of the model as a function of the decay exponent, $\alpha$. The flow equations reveal an attractive fixed point for the distribution 
of the hoppings at $\alpha=0$, which corresponds to the GOE phase. Rather surprisingly, we find that $\alpha=\frac{1}{2}$ is an unstable fixed point and 
for $\alpha>\frac{1}{2}$, the distribution of hoppings remains fixed under the flow. The strong-bond RG procedure inspired from the flow equations  provides
an intuitive description of the emergence of this transition in the level-spacing statistics. The signature we use to probe the localization transition at $\alpha =1$ is in the distribution
of the level spacings.

The results discussed in this paper can be generalized to study other
systems. The particular advantage of this method is that it preserves
the \emph{full spectrum} of the Hamiltonian. This has implications
in studying localization-delocalization transitions in interacting
and disordered systems. Many-body localized systems are pseudo-integrable,
in the sense that they have a large number of conserved charges with
local support \cite{SerbynPRL2013,HuseNandkishorePhenom2014}. There
has been some recent work on studying these conserved quantities using
various methods (for instance, see Ref.~\onlinecite{RademakerPRL}). These conserved quantities can be obtained directly using flow
equations, and therefore this method provides a tool to
study fully-localized interacting phases, as recently reported in Ref.~{\onlinecite{pekker2016Wegner}}.

In this paper, we have also shown that the strong-bond RG procedure
is suitable to study critically \emph{delocalized} phases. We expect
that a similar generalized method should be useful to study the system
across the MBL-ergodic phase transition. 

Yet another direction to consider is the analytical description of the phases of disordered and interacting systems with power-law decays. In these systems, the strong-disorder renormalization techniques developed so far fails, and the only known results are obtained numerically via exact diagonalization \cite{GarciaGarciaPRE2012}. A strong-disorder 
renormalization group suitable to handle such systems
is still missing. We expect the flow equation technique to be more
useful in that task, to study the both zero- and high-temperature
phases. \textbf{}%
\begin{comment}
\textbf{{[}why?{]}} 
\end{comment}

\section{{\normalsize{}Acknowledgments}}

The authors would like to acknowledge useful discussions with Stefan
Kehrein, Sarang Gopalakrishnan, Eduardo Miranda and David Huse. P.T.
and G. R. are grateful for support from NSF through DMR-1410435, as
well as the Institute of Quantum Information and matter, an NSF Frontier
center funded by the Gordon and Betty Moore Foundation, and the Packard
Foundation. V.L.Q. acknowledges financial support from FAPESP, through
grants 2012/17082-7 and 2009/17531-3.\bigskip{}

\appendix

\section{{\normalsize{} Critical point analysis 
\label{sec:Appendix-Levitovs-analysis}}}

In this appendix, we outline the analysis of the the transition as
a function of the power law exponent $\alpha$. The results here are an extension of Levitov`s results for the $\alpha=1$ critical point \cite{Levitov2,LevitovPRL}.

\selectlanguage{american}%
Consider an arbitrary
site at $r_{i}$ and two concentric one-dimensional ``spheres'' with radius 
$2^{k}R<\left|r-r_{i}\right|<2^{k+1}R$,
for a given value of $k$. The volume of this shell is 
$V\left(k,R\right)=2^{k}R$.
The characteristic level spacing in this shell, $\Delta\left(k,R\right)$,
and the typical hopping strength, $J\left(k,R\right)$, are 

\begin{eqnarray}
\Delta\left(k,R\right) & \sim & \frac{1}{nV\left(k,R\right)},\\
J\left(k,R\right) & \sim & \frac{1}{\left(2^{k}R\right)^{\alpha}},
\end{eqnarray}
where we have assumed a constant density of states $n$. The typical
value of the probability distribution for a resonance with site $i$
in this shell is 

\begin{equation}
P\left(k,R\right)\propto\frac{J\left(k,R\right)}{\Delta\left(k,R\right)}=\frac{
2^{k}R}{\left(2^{k}R\right)^{\alpha}}=\left(2^{k}R\right)^{\epsilon},
\end{equation}
where we defined $\epsilon=1-\alpha$. Notice that the volume of the
system is $V_{tot}=2^{L}R$. There are three possible cases that
must be considered separately. In all cases, we consider additionally the probability
of \textit{not} finding a resonance beyond a $R$, $P_{nr}$. Delocalized
phases, as well as the critical point, will have a vanishing $P_{nr}$
for large $R$.

\begin{itemize}
\item Critical regime, $\epsilon=0$: $P\left(k,R\right)=b$ is a constant.
The probability of not finding a resonance beyond a $R$ is 
$P_{nr}=\prod_{k=1}^{N}\left[1-P\left(k,R\right)\right]=\left(1-b\right){}^{L}
\rightarrow0$
at the thermodynamic limit. Also, the total number of sites in resonance
with site $i$ is 
\end{itemize}
\begin{eqnarray}
N_{res} & = & \sum_{k=0}^{N}P\left(k,R\right)=\left(L+1\right)b\nonumber \\
 & \sim & \log\left(V_{tot}\right).
\end{eqnarray}

\begin{itemize}
\item Delocalized regime, $\epsilon>0$: In this case, we take the log to find
\end{itemize}
\begin{eqnarray}
\log\left(P_{nr}\right) & = & 
\sum_{k=0}^{N}\log\left(1-\left(2^{k}R\right)^{\epsilon}\right)\nonumber \\
 & = & 
-\sum_{k=0}^{N}\frac{1}{k}\left(2^{k}R\right)^{\epsilon}\sim-R^{\epsilon}2^{
\epsilon L}\\
\Rightarrow & P_{nr} & \sim\exp\left(-R^{\epsilon}2^{\epsilon 
L}\right)\rightarrow0
\end{eqnarray}
Note that, in general, the probability of a resonance 
$P\left(k,R\right)=1-P_{rn}$
grows with $R$ for $\epsilon>0$ indicating a delocalized regime.
In fact, the number of sites at resonance is,

\begin{eqnarray}
N_{res} & = & 
\sum_{k=0}^{N}\left(2^{k}R\right)^{\epsilon}=R^{\epsilon}\sum_{k=0}^{N}\left(2^{
\epsilon}\right)^{k}\nonumber \\
 & = & 
R^{\epsilon}\frac{\left(1-2^{\epsilon\left(N+1\right)}\right)}{1-2^{\epsilon}}
\sim R^{\epsilon}2^{\epsilon N}\text{ for }N\gg\frac{1}{\epsilon}\nonumber \\
 & \propto & \left(V_{tot}\right)^{\epsilon},
\end{eqnarray}
which diverges at the thermodynamic limit. It should also be noticed
that such divergence is not extensive in volume, but instead increases
with power $\epsilon$.
\begin{itemize}
\item Localized Regime, $\epsilon<0$: Similar to the delocalized regime,
we set $\epsilon=0^{-}$, then 
$P_{nr}=\prod_{k=0}^{N}\left(1-\left(2^{k}R\right)^{\epsilon}
\right)\sim\exp\left(R^{\epsilon}\right)$.
So for large enough $R$, $P_{nr}\rightarrow1$, indicating a localized
phase. Equivalently, if the number of sites in resonance, 
\end{itemize}
\begin{equation}
N_{\textrm{res}}\sim\mbox{const}.
\end{equation}

which also points to the fact that the phase is localized, since $N_{\textrm{res}}$
does not scale with system size.

\section{{\normalsize{}Simple Example: Spin-1/2 in Magnetic field 
\label{sec:Appendix-Simple-Example-Spin-1/2}}}

As a simple example, we use the FET to diagonalize the Hamiltonian
describing a spin-$\frac{1}{2}$ particle in a magnetic field. Considering
a magnetic field parallel to the $xz$ plane, and call its components $J$ and 
$h$, to keep the analogy with the main text. The Hamiltonian is

\begin{equation}
H=h\sigma_{z}+J\sigma_{x},\label{eq:example-Hamil}
\end{equation}
where we choose $J$ and $h$ such that $\sqrt{J^{2}+h^{2}}=1$. This Hamiltonian
is $2\times2$ matrix which can be easily diagonalized to obtain
the eigenvalues $\pm1$. We now solve this eigenvalue problem
using the FET. Defining $H_{0}=h\sigma_{z}$ and $V=J\sigma_{x}$, the
generator is given by

\begin{eqnarray}
\eta & = & \left[H_{0},V\right]=i2hJ\sigma_{y}.\label{eq:example-generator}
\end{eqnarray}
The equation of motion for $H\left(\Gamma\right)$ (see Eq. 
(\ref{eq:HamiltonianRGTimeevolution}))
becomes,
\begin{eqnarray}
\frac{dH}{d\Gamma} & = & \left[\eta,H\right]\nonumber \\
 & =- & 4hJ\left(h\sigma_{x}-J\sigma_{z}\right)\label{eq:example-eqofmotion}
\end{eqnarray}

Using Eqs.~(\ref{eq:example-Hamil}) and (\ref{eq:example-eqofmotion}),
we find
\begin{eqnarray}
\frac{d}{d\Gamma}h\left(\Gamma\right) & = & 
4h\left(\Gamma\right)J\left(\Gamma\right)^{2}\label{eq:example-flow_h}\\
\frac{d}{d\Gamma}J\left(\Gamma\right) & = & 
-4h\left(\Gamma\right)^{2}J\left(\Gamma\right)\label{eq:example-flow_J}
\end{eqnarray}
with the initial conditions, $h\left(0\right)=h$, and $J\left(0\right)=J$.
We note the above equations have a conserved quantity, 
$h^{2}\left(\Gamma\right)+J^{2}\left(\Gamma\right)={\rm const.=}1$,
which describes a circle in the parameter space. Parametrization in
terms of trigonometric functions, 
$h\left(\Gamma\right)=\cos\theta\left(\Gamma\right)$,
and $J\left(\Gamma\right)=\sin\theta\left(\Gamma\right)$, transforms
the the problem into a single-variable equation,

\begin{eqnarray}
\frac{d}{d\Gamma}\theta\left(\Gamma\right) & = & 
-2\sin\left(2\theta\left(\Gamma\right)\right),\label{eq:example-flow_theta}
\end{eqnarray}
which gives the solution,
\begin{eqnarray}
\theta\left(\Gamma\right) & =\arctan & \left(\frac{J}{h}e^{-4\Gamma}\right).
\end{eqnarray}
In the limit $\Gamma\rightarrow\infty$, the parametric 
angle $\theta\left(\Gamma\right)$
vanishes, which implies $h\left(\infty\right)=1$ and $J\left(\infty\right)=0$,
so that the Hamiltonian is diagonal and the eigenvalues are $\pm1$.
Another equivalent way of finding the eigenvalues is using the unitary
transformation explicitly 

\begin{eqnarray}
H\left(\Gamma\right) = 
e^{\int_{0}^{\Gamma}d\Gamma'\eta}H\left(0\right)e^{-\int_{0}^{\Gamma}
d\Gamma'\eta},
\end{eqnarray}
with
\begin{equation}
\int d\Gamma'\ \eta\left(\Gamma'\right) =
-i\sigma_{y}\frac{1}{2}\arctan\left(\frac{J}{h}\right),
\end{equation}

where the rotation operator is $S_{y}=\frac{1}{2}\sigma_{y}$ and the rotation angle is $\theta=\arctan\left(\frac{J}{h}\right)$.
This is exactly the rotation that diagonalizes the Hamiltonian

\begin{equation}
e^{-i\theta S_{y}}\left(h\sigma_{z}+J\sigma_{x}\right)e^{i\theta 
S_{y}}=\sigma_{z}.
\end{equation}

This simple example illustrates the basic steps of how to implement
the FET for a generic Hamiltonian. The first step is to split $H$
into $H_{0}$ and $V$ such that 
$\mbox{Tr}\left(\frac{dH_{0}}{d\Gamma}V\right)=0$.
After that, the computation of $\eta=\left[H_{0},V\right]$ and the
flow equations is straightforward algebra, except when extra terms
are generated. In general, it is not possible to solve the flow equations,
but in this case the solution is simple, showing the exponential decay
in $\Gamma$ of the off-diagonal operator $V$.

\section{{\normalsize{}Two-site solution: Details 
\label{sub:Appendix-Two-site-solution}}}

\begin{figure}
\selectlanguage{english}%
\includegraphics[width=1\columnwidth]{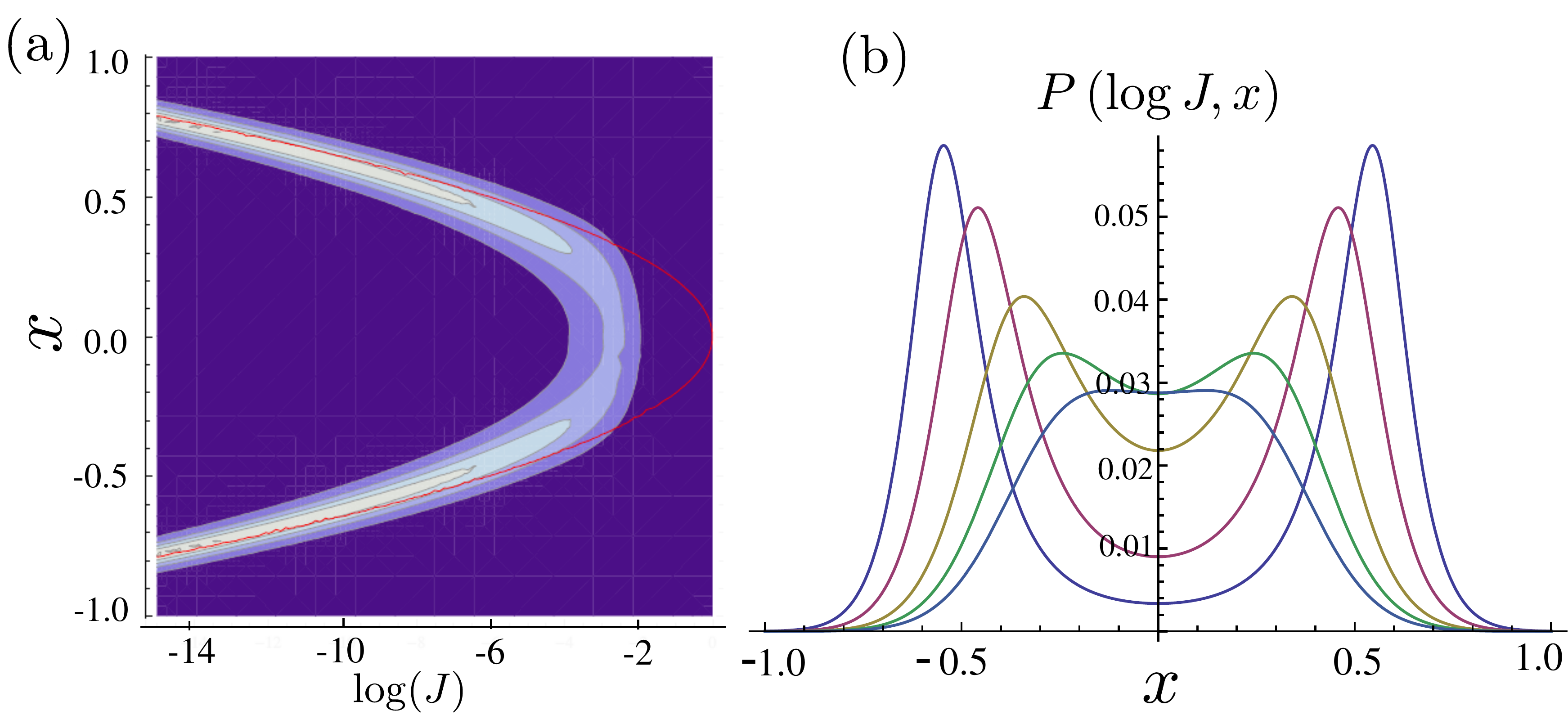}

\selectlanguage{american}%
\caption{\selectlanguage{english}%
(Color Online) (a) Density plot of the distribution $x$ and $\log J$ at $\Gamma/4=6$.
In red, \foreignlanguage{american}{the curve $x^{2}\Gamma=-\log\left(J\right)$,
showing the maximum intensity for small $\log J$.} (b) Cuts of $\log J=-5$
  (blue), $-4$ (purple), $-3$ (yellow), $-2.5$ (green), $-2.2$ (blue)
at time $\frac{\Gamma}{4}=\frac{10}{3}$. The two distinct peaks collapse
at large values of $\log J$. 
\label{fig:appendix-twositecorrelationplot}\selectlanguage{american}%
}
\end{figure}

The exact solution to the two-site FE defined in 
Section~\ref{sec:Method:-Wegners-Flow} is

\begin{equation}
J=J_{0}\frac{\exp\left(-\frac{r^{2}}{2}\Gamma\right)\sqrt{\left(2J_{0}\right)^{2
}+x_{0}^{2}}}{\sqrt{\left(2J_{0}\right)^{2}\exp\left(-r^{2}\Gamma\right)+x_{0}^{
2}\exp\left(r^{2}\Gamma\right)}},
\end{equation}

\begin{equation}
x=x_{0}\frac{\exp\left(\frac{r^{2}}{2}\Gamma\right)\sqrt{\left(2J_{0}\right)^{2}
+x_{0}^{2}}}{\sqrt{\left(2J_{0}\right)^{2}\exp\left(-r^{2}\Gamma\right)+x_{0}^{2
}\exp\left(r^{2}\Gamma\right)}},
\end{equation}
where we have chosen the initial conditions, 
$\left(J\left(\Gamma=0\right),x\left(\Gamma=0\right)\right)\equiv\left(J_{0},x_{
0}\right)$.
As noted in Section~\ref{sec:Method:-Wegners-Flow}, it is convenient
to change variables to $\left(r,\theta\right)$, where $r^{2}=4J^{2}+x^{2}$,
and $\tan\theta=2J/x$. The solution for the flow of distributions
is easily obtained, 
$\tilde{P}\left(\tan\theta\left(\Gamma\right),r\left(\Gamma\right)\right)=\tilde
{P}\left(\tan\theta_{0},r_{0}\right)\exp\left(r_{0}^{2}\Gamma\right)$,
where $\tilde{P}\left(\tan\theta_{0},r_{o}\right)$ is the initial distribution
of $\tan\theta\left(\Gamma\right)$ and $r\left(\Gamma\right)$. Consequently,
the distribution of the variables, $\left(J,x\right)$, is obtained
using the Jacobian of the $\left(\tan\theta,r\right)\rightarrow\left(J,x\right)$
transformation, 

\begin{equation}
P\left(J,x\right)=P\left(J_{0},x_{0}\right)\frac{\left(x^{2}+4J^{2}
\right)\exp\left[\left(4J^{2}+x^{2}\right)\Gamma\right]}{\left(x^{2}+4J^{2}
\exp\left[2\left(4J^{2}+x^{2}\right)\Gamma\right]\right)}.
\end{equation}
In the long-time limit, the distribution becomes

\begin{equation}
P\left(J,x\right)\approx 
P_{1}\left(\log\left(J_{0}\right),x_{0}\right)\frac{x^{2}
\exp\left(\log\left(J\right)+x^{2}\Gamma\right)}{4\exp\left[
2\left(\log\left(J\right)+x^{2}\Gamma\right)\right]+x^{2}},
\end{equation}
where the maximum for the surface plot of $P\left(J,x\right)$ is
at the curve $x^{2}\Gamma=-\log\left(J\right)$. In Fig.~\ref{fig:appendix-twositecorrelationplot},
we plot the correlation between $\log J\left(\Gamma\right)$ and 
$x\left(\Gamma\right)$
at $\Gamma/4=6$. The initial couplings $\left(J_{0},x_{0}\right)$
were chosen with $J_{0}$ uniformly distributed between $\left[0,1\right]$
and the fields on the two sites, $h_{1}$ and $h_{2}$ are also uniformly
distributed between $\left[0,1\right]$.

\section{{\normalsize{}Initial Distribution of hoppings 
\label{sec:Initial_hoppings}}}

\begin{figure}
\includegraphics[width=1\columnwidth]{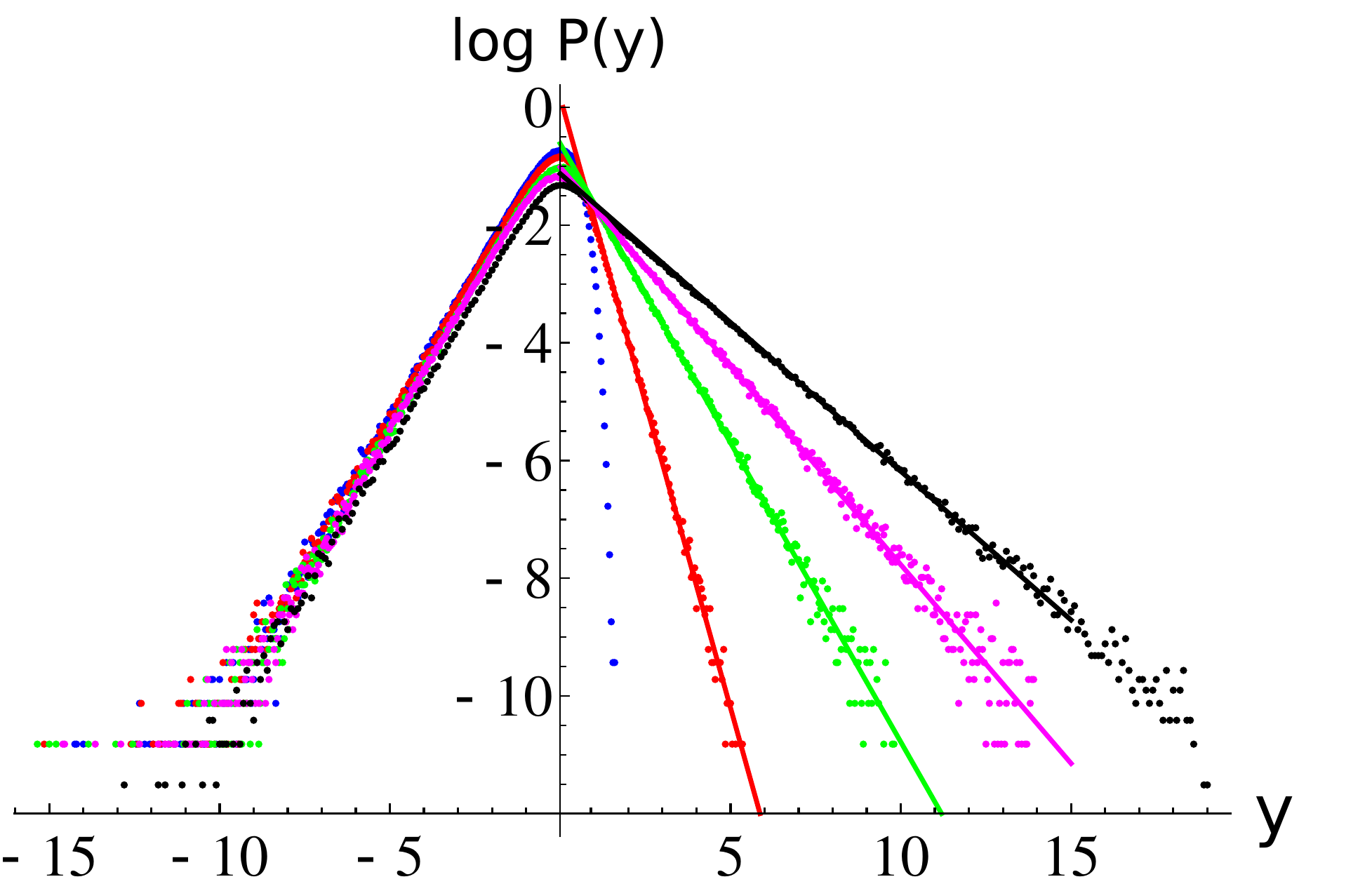}

\caption{(Color Online) Initial probability distribution $ P\left(y=\log{J}\right)$, of couplings
connected to an arbitrary test site, in log scale. The distributions
have been shifted horizontally so that the maximum of all the curves
are located at $y=0$. For $y<0$, the uniform part of the distribution
$P\left(J\right)$, corresponding to $\log P\left(y\right)\sim-y$, is independent of $\alpha$. For $y>0$, the angular coefficient, expected to result in $-\frac{1}{\alpha}$,
gives $-2.07$ for $\alpha=\frac{1}{2}$ (red line and points), $-1.01$ for $\alpha=1$ (purple line and points), and $-0.51$ for $\alpha=2$  (black line and points). The blue points correspond
to $\alpha=0$ , where the saddle point approximation fails. \label{fig:init-probability-distribution}}

\end{figure}

In this Appendix, we derive the form of the initial marginal distribution
for all hoppings, $P_{\Gamma=0}\left(J\right)$. The distribution of
all bonds connected to single site, say $i$, is

\begin{equation}
P_{\Gamma=0}\left(J\right)=\frac{1}{N}\sum_{j=1}^{N-1}P_{\left|i-j\right|}\left(J_{i}^{j}
\right)\delta\left(J-J_{i}^{j}\right),\label{eq:discretesumbonds}
\end{equation}
where $P_{\left|i-j\right|}\left(J_{i}^{j}\right)$ corresponds to
the distribution of bonds of a particular length. Note that the left-hand
side is independent of $i$. We restrict ourselves to the case where
each of the bonds are normally distributed. This assumption is sufficient since,
as we have shown in Section~\ref{sec:Flow-RG}, all initial distributions
of scale-invariant hoppings $G_{i}^{j}$ flow to normal distributions under the RG
procedure. Setting $l=\left|i-j\right|$ we have

\begin{equation}
P_{l}\left(J_{i}^{j}\right)=\frac{1}{\sqrt{2\pi\sigma_{l}^{2}}}\exp\left(-\frac{
\left(J_{i}^{j}\right)^{2}}{2\sigma_{l}^{2}}\right),
\end{equation}
where $\sigma_{l}=\sigma_{0}/l^{\alpha}$.  Now we evaluate the approximate form
of $P\left(J\right)$ by taking the continuum limit and setting $\sigma_{0}=1$.

\begin{eqnarray}
P\left(J\right) & = & 
\frac{1}{N}\int_{0}^{N}dx\frac{1}{\sqrt{2\pi}}\exp\left(-\frac{J^{2}x^{2\alpha}}
{2}\right)x^{\alpha},\nonumber \\
 & = & 
\frac{1}{N\sqrt{2\pi}}\int_{0}^{N}dx\exp\left(-\frac{J^{2}x^{2\alpha}}{2}
+\alpha\ln x\right).\label{eq:marginalP(J)saddlepoint}
\end{eqnarray}
We can evaluate Eq.~(\ref{eq:marginalP(J)saddlepoint}) using the
saddle point approximation. The saddle point for the function 
$f\left(x\right)=-\frac{J^{2}x^{2\alpha}}{2}+\alpha\ln x$
is given by the condition $f'\left(x^{*}\right)=0$, that is, $x^{*}=J^{-1/\alpha}$.
This is a maximum as evidenced by 
$f''\left(x^{*}=J^{-1/\alpha}\right)=-2\alpha^{2}/x^{2}<0$.
Now, evaluating Eq.~(\ref{eq:marginalP(J)saddlepoint}) by expanding
around the saddle point, we obtain,

\begin{eqnarray}
P\left(J\right) & = & 
\frac{\exp\left(-\frac{1}{2}\right)}{N\sqrt{2\pi}J}\int_{0}^{N}dx\exp\left(-2J^{
\frac{2}{\alpha}}\alpha^{2}\left(x-J^{-\frac{1}{\alpha}}\right)^{2}
\right)\nonumber \\
 & \sim & 
\frac{\exp\left(-\frac{1}{2}\right)}{4N\alpha\sqrt{2\pi}J^{1+\frac{1}{\alpha}}},
\end{eqnarray}
 where in the second step we used the limit of large $N$, to approximate
$\text{{erf}}\left(N\alpha\sqrt{2}J^{\frac{1}{\alpha}}\right)\approx1$.
Ultimately, the distribution of the bonds becomes

\begin{equation}
P\left(J\right)=\frac{C_{\alpha}}{J^{1+\frac{1}{\alpha}}}.\label{
eq:marginalJdist1+1/a}
\end{equation}
The validity of the saddle point introduces a finite-size cutoff,
dependent on the system size $N$. For the saddle-point approximation
to be valid, we require $J^{-1/\alpha}<N$, which means it fails for
$J<J_{{\rm c}}\equiv\frac{1}{N^{\alpha}}$. The bonds below $J_{{\rm 
c}}$
are set to $J=0$. The distribution becomes uniformly distributed since

\begin{eqnarray}
P\left(J\right) & \approx & 
\frac{1}{N}\int_{0}^{N}dx\frac{1}{\sqrt{2\pi}}x^{\alpha},\nonumber \\
 & = & 
\frac{N^{\alpha}}{\left(\alpha+1\right)\sqrt{2\pi}}.\label{
eq:uniformdistbelowcutoff}
\end{eqnarray}
An example of the distribution $P\left(J\right)$ is given in 
Fig.\ref{fig:init-probability-distribution},
where we consider a site connected to 100 neighbors (average over
70 realizations). Working in log scale, we find the following behavior
of $\log P\left(y=\log{x}\right)$,

\begin{equation}
\log P\left(y\right)\sim\begin{cases}
y & ,y<0\\
-\frac{y}{\alpha} & ,y>0
\end{cases},
\end{equation}
where we have also shifted the distribution (in log scale) so that the crossover
point is at $y=0$.

We note that the calculation done here is approximate. The distribution
we find is clearly incorrect in the limits of $\alpha\rightarrow0$ and 
$\alpha\rightarrow\infty$.
In the limit $\alpha\rightarrow0$, the saddle point calculation is
not trustworthy, while in the limit $\alpha\rightarrow\infty$ the
continuum approximation done to Eq.~(\ref{eq:discretesumbonds})
is no longer valid.

\section{{\normalsize{}Effect of hoppings on bandwidth 
\label{sec:app-bandwidth}}}

In light of the fact that the power law exponent of the distribution
of $J_{i}^{j}$ does not flow for $\alpha>1/2$, we rewrite its evolution as 

\begin{equation}
J_{i}^{j}\left(\Gamma\right)=G_{i}^{j}\frac{f_{\Gamma}\left(x_{i}^{j}\right)}{
\left|i-j\right|^{\alpha}},
\end{equation}
where $G_{i}^{j}$ is a scale-invariant random number and $f_{\Gamma}\left(x_{i}^{j}\right)$
takes into account effects of the $x$ variables into the $J$ evolution.
At the starting point, $f_{\Gamma=0}\left(x_{i}^{j}\right)=1$. The
equation for the evolution of $f_{\Gamma}\left(x\right)$ follows
from Eq.~(\ref{eq:h_evol}),

\begin{equation}
\frac{df_{\Gamma}}{d\Gamma}\left(x\right)=-x^{2}f_{\Gamma}\left(x\right)
\end{equation}
whose solution is $f_{\Gamma}\left(x\right)=e^{-\Gamma x^{2}}$. Integrating the evolution of $h_{i}$,  Eq.~(\ref{eq:h_evol}), we compute the typical field change $\Delta 
h_{i}=h_{i}\left(\infty\right)-h_{i}\left(0\right)$,
that summarizes the effects of hoppings in the field evolution, and therefore gives the bandwidth

\begin{eqnarray}
\Delta h_{i} & \approx & 
N^{1-2\alpha}\frac{1}{2\Gamma}\int^{\frac{1}{N^{\alpha}}}d\Gamma e^{-\Gamma x^{2}}\\
 & \sim & N^{1-2\alpha}\log N.
\end{eqnarray}
If $\alpha\le\frac{1}{2}$, the bandwidth diverges in the thermodynamic limit, while at $\alpha>\frac{1}{2}$
it stays of $\mathcal{O}\left(1\right)$. The logarithmic correction for $\alpha=1/2$ indicates a critical behavior. This is the result quoted in the main text.

\selectlanguage{english}%

\section{{\normalsize{}Strong-bond RG: 
Details\label{sec:Appendix-Flow-RG:-Derivation}}}

\subsection{RG step derivation.}

In this Appendix, we give an outline of some of the derivations used in the main text, in the Strong-bond
RG procedure. We start deriving 
Eqs.~(\ref{eq:FlowRG-Jrenorm-1})-(\ref{eq:FlowRG-hrenorm}).
To abbreviate the notation, let us define $c_{i}^{\dagger}\equiv \tilde{i}$ and $c_{i}\equiv i$. We consider the two-site chain, since
this is the building block for the RG steps. The idea is to solve
this chain in leading order of $\frac{1}{\Gamma}$. The Hamiltonian
for three sites is given by

\begin{eqnarray}
H_{3s} & = & 
J_{1}^{2}\left(\tilde{1}2+\tilde{2}1\right)+J_{2}^{3}\left(\tilde{2}3+\tilde{3}
2\right)+J_{1}^{3}\left(\tilde{1}3+\tilde{3}1\right)\nonumber \\
 & + & h_{1}\tilde{1}1+h_{2}\tilde{2}2+h_{3}\tilde{3}3
\end{eqnarray}

Calculating the generator explicitly, we find

\begin{eqnarray}
\eta & = & 
J_{1}^{2}\left(h_{1}-h_{2}\right)\left(\tilde{2}1-\tilde{1}2\right)+J_{2}^{3}
\left(h_{2}-h_{3}\right)\left(\tilde{3}2-\tilde{2}3\right)\nonumber \\
 & + & J_{1}^{3}\left(h_{1}-h_{3}\right)\left(\tilde{3}1-\tilde{1}3\right)
\end{eqnarray}

We now make the assumption that $r_{12}\gg r_{23}$. Under this assumption, we consider the evolution in an interval from  from $\Gamma=0$ to $\delta\Gamma\sim1/r_{12}^2$  where only bonds and fields related to sites 1 and 2 evolve, while the couplings in other sites change infinitesimally. Keeping the leading order terms (zeroth order in $\delta\Gamma$), we find the evolution of $\eta$ to be 

\begin{eqnarray}
\int d\Gamma\eta & = & \int d\Gamma 
J_{1}^{2}\left(h_{1}-h_{2}\right)\left(\tilde{2}1-\tilde{1}2\right)+\mathcal{O}
\left(\delta\Gamma\right)\nonumber \\
 & = & 
\alpha_{12}\left(\tilde{2}1-\tilde{1}2\right)+\mathcal{O}
\left(\delta\Gamma\right)
\end{eqnarray}
where 
$\alpha_{12}$ was defined in Eq. (\ref{eq:maxrbondangle}).
Higher order corrections can be neglected under the assumption that $\delta\Gamma$
is sufficiently small. Defining $A=\int d\Gamma\eta$ and recalling
the Baker-Campbell-Hausdorff formula,

\begin{eqnarray}
e^{A}He^{-A} & = & 
H+\left[A,H\right]+\frac{1}{2}\left[A,\left[A,H\right]\right]+\ldots
\end{eqnarray}
we find the leading order correction to be

\begin{eqnarray}
\left[H,\alpha_{12}\left(\tilde{2}1-\tilde{1}2\right)\right] & = & 
J_{2}^{3}\alpha_{12}\left(\tilde{3}1+1\tilde{3}\right)-J_{1}^{3}\alpha_{12}
\left(3\tilde{2}+2\tilde{3}\right).
\end{eqnarray}

Therefore, the commutators with the Hamiltonian yield

\begin{eqnarray}
\left[A,H\right] & = & 
J_{1}^{3}\alpha_{12}\left(3\tilde{2}+2\tilde{3}\right)-J_{2}^{3}\alpha_{12}
\left(\tilde{3}1+1\tilde{3}\right),\\
\left[A,\left[A,H\right]\right] & = & 
-J_{2}^{3}\alpha_{12}\left(3\tilde{2}+2\tilde{3}\right)-J_{1}^{2}\alpha_{12}
\left(\tilde{3}1+1\tilde{3}\right).
\end{eqnarray}

One can easily show, by induction, that summing the series, we find

\begin{eqnarray}
H\left(\delta\Gamma\right) & = & 
\left(J_{1}^{3}\cos\alpha_{12}+J_{2}^{3}\sin\alpha_{12}\right)\left(\tilde{3}
1+1\tilde{3}\right)\nonumber \\
 & + & 
\left(J_{2}^{3}\cos\alpha_{12}-J_{1}^{3}\sin\alpha_{12}\right)\left(\tilde{3}
2+2\tilde{3}\right).
\end{eqnarray}
and this is the basis for the RG equations (\ref{eq:FlowRG-Jrenorm-1}) and
(\ref{eq:FlowRG-Jrenorm-2}). The change in the fields, Eq.~(\ref{eq:FlowRG-hrenorm}), can be found with similar reasoning. The generalization for chain of $N$ sites is trivial, since
the RG procedure can be thought as processes acting on blocks of three
sites.

\subsection{Additional numerical results}

\begin{figure}
\includegraphics[width=1\columnwidth]{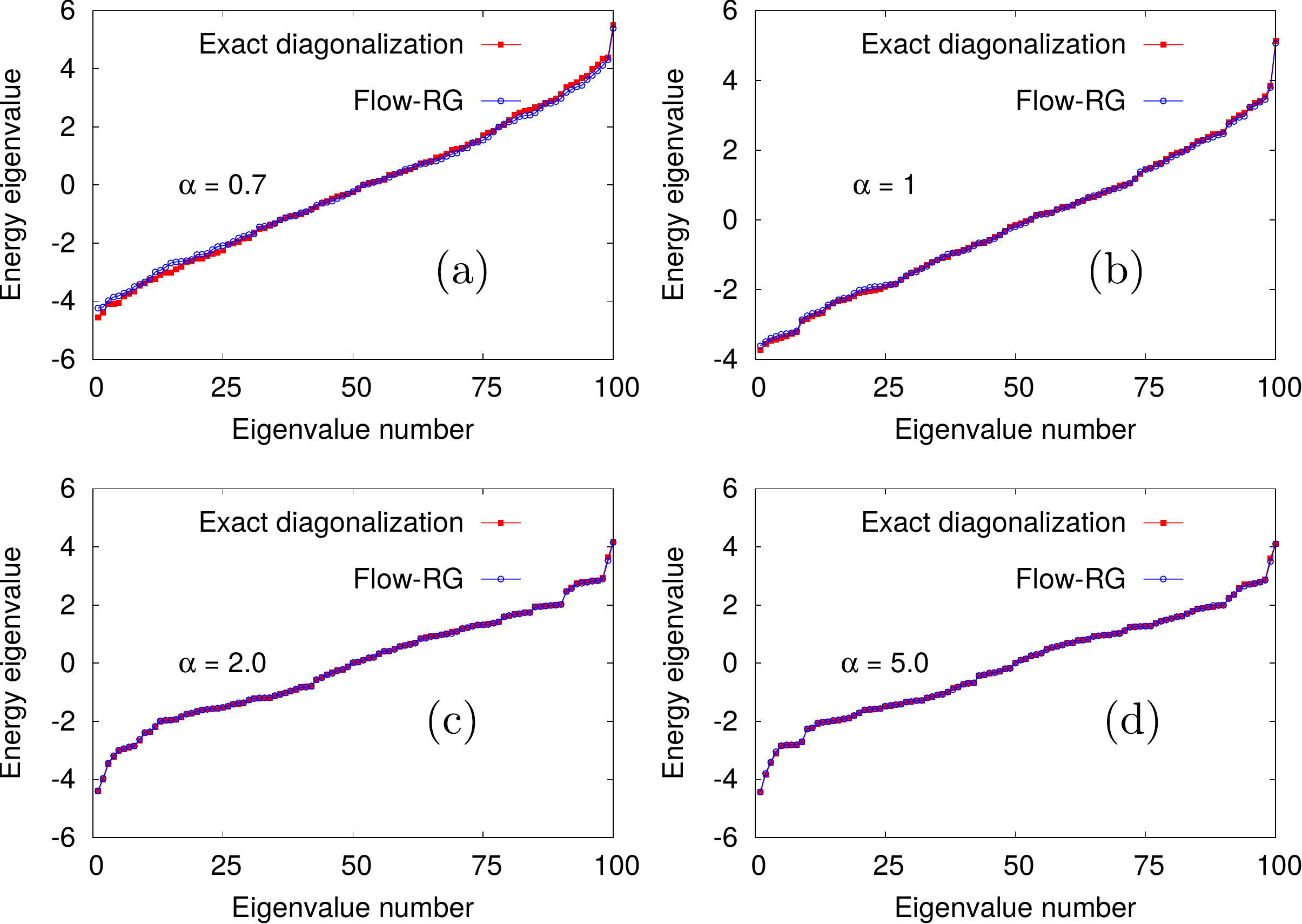}

\caption{Comparison of the single particle spectrum obtained from 
exact-diagonalization
with the one obtained from the Strong-bond RG technique. (a)-(d) in ascending order of exponents, $\alpha=0,$ $0.7$, $2.0$, $5.0$. In all cases, both spectra look reasonably similar.
A careful inspection of the level-spacing statistics, however, reveals
that the eigenvalues obtained in case (a), $\alpha=0$, does not experience
repulsion, like a delocalized phase should (see main text for further details about the level spacing). \label{fig:FlowRGvsEDcomparison}}
\end{figure}

Numerically, we can compare the spectrum obtained from exact diagonalization
and the RG procedure. In Fig.~\ref{fig:FlowRGvsEDcomparison}, we
compare all the single-particle levels from the two methods, obtained for a
single disorder realization in a chain of $N=100$ sites. We plot for exponents $\alpha=0.7$,
$1.0$, $2.0$ and $5.0$, with Gaussian distribution with unit standard deviation of $h_{i}$ and $G_{i}^{j}$. Clearly we obtain very good agreement between the two procedures in all the cases. The level spacing is more subtle, and studied in the main text. 

\section{{\normalsize{}Master Equation\label{sec:Master-Equation}}}

\textbf{}%
\begin{comment}
\textbf{{[}Victor, Please check the master equation section. I have
modified the appendix to reflect our method.{]}}
\end{comment}

In this appendix, we write down the master equation for the distribution
of $J$, $ P_{\Gamma}\left(J\right)$, under the RG procedure. We consider the shell in the $J-x$ phase space
with radius $r=\sqrt{x^{2}+4J^{2}}$ and width $dr$. We denote
the set of all hopping terms by a single variable $J$, and field
differences by $x=\left|h_{i}-h_{j}\right|$. Let the distribution
of the $x$ in the shell be $n\left(x=r\right)$ and the distribution
of the bonds be $P_{\Gamma}\left(J\right)$. 

Consider the effects of the decimation of a pair
with hopping $J_{i}^{j}$. The $J$ distribution changes due to the
removal of a large hopping $J_{i}^{j}$ and by the renormalization
of all couplings connected to sites $i$ or $j$. The change in the
distribution of the bonds, $\Delta P_{\Gamma}\left(J\right)$, is
\begin{widetext}
\begin{eqnarray}
\Delta P_{\Gamma}\left(J\right)=dr\int_{0}^{2\pi}d\theta_{i}^{j}\int 
dxdJ_{i}^{j}n\left(x\right)P_{\Gamma}\left(J_{i}^{j}\right)\delta\left(\sqrt{\left(2 J_{i}^
{j}\right)^{2}+\left(x_{i}^{j}\right)^{2}}-r\right)\int\prod_{k}\left(dJ_{k}^{i}
dJ_{k}^{j}\right)P_{\Gamma}\left(J_{k}^{i}\right)P_{\Gamma}\left(J_{k}^{j}\right)\times\nonumber 
\\
\times\sum_{k}\left[\delta\left(J-\tilde{J}_{k}^{i}\right)+\delta\left(J-\tilde{
J}_{k}^{j}\right)-\delta\left(J-J_{k}^{i}\right)-\delta\left(J-J_{k}^{j}
\right)-\delta\left(J-J_{i}^{j}\right)\right],
\end{eqnarray}
\end{widetext}
where $\tilde{J}$ variables are defined in the main text, Eqs. 
(\ref{eq:FlowRG-Jrenorm-1})
and (\ref{eq:FlowRG-Jrenorm-2}) and the angle variable is defined in Eq.(~\ref{eq:thetadefinition}).
Let us now define 
\begin{equation}
\eta\left(r,J_{i}^{j}\right)=\int 
dxn\left(x\right)\delta\left(\sqrt{\left(2J_{i}^{j}\right)^{2}+x^{2}}-r\right).
\end{equation}

We make a 
simplification $\eta\left(r,J_{i}^{j}\right)\approx\eta\left(r\right)\approx\eta$
,
where $\eta$ a constant. This approximation relies on the fact that,
for most of the bonds, $J_{i}^{j}\ll x_{i}^{j}$ and so, we approximate
$J_{i}^{j}\approx0$. As the largest $r$-bonds are removed from the
chain, the normalization also changes. The number of removed couplings
is $\eta dr$ and an overall pre-factor $\left(1-\eta dr\right)^{-1}$
must be included. Therefore, the new distribution $\tilde{P}\left(J\right)$
is given by a sum of the previous distribution $P\left(J\right)$
and the above contribution $\Delta P\left(J\right)$, multiplied by the normalization pre-factor,

\begin{widetext} 

\begin{eqnarray}
\tilde{P}\left(J\right) & = & \frac{1}{1-\eta dr}\left[P\left(J\right)+\eta 
dr\int d\theta_{i}^{j}\int 
dJ_{i}^{j}P\left(J_{i}^{j}\right)\int\prod_{k}\left(dJ_{k}^{i}dJ_{k}^{j}
\right)\times\right.\nonumber \\
 & \times & 
\left.P\left(J_{k}^{i}\right)P\left(J_{k}^{j}\right)\left(\sum_{k,p=i,j}
\delta\left(J-\tilde{J}_{k}^{p}\right)+\delta\left(J-J_{k}^{p}
\right)-\delta\left(J-J_{i}^{j}\right)\right)\right]\\
\implies\tilde{P}\left(J\right) & = & P\left(J\right)+\eta dr\int 
dJ_{i}^{j}P\left(J_{i}^{j}\right)\int\prod_{k}dJ_{k}^{i}dJ_{k}^{j}P\left(J_{k}^{
i}\right)P\left(J_{k}^{j}\right)\sum_{k}\left(\sum_{p=\left\{ i,j\right\} 
}\delta\left(J-\tilde{J}_{k}^{p}\right)-\delta\left(J-J_{k}^{p}
\right)\right)\nonumber 
\end{eqnarray}

\end{widetext}

For simplicity we assume $n\left(r\right)$ to be a constant. We now
take the continuum limit as the scale $r$ reduces. The master equation
becomes

\begin{widetext}

\[
\frac{\partial P\left(J\right)}{\partial r}=\eta\int d\theta_{i}^{j}\int 
dJ_{i}^{j}P\left(J_{i}^{j}\right)\int\prod_{k}dJ_{k}^{i}dJ_{k}^{j}P\left(J_{k}^{
i}\right)P\left(J_{k}^{j}\right)\sum_{k}\left(\sum_{p=\left\{ i,j\right\} 
}\delta\left(J-\tilde{J}_{k}^{p}\right)-\delta\left(J-J_{k}^{p}\right)\right).
\]

Notice that this equation keeps track of how the distribution changes with $r$, not with the scale $\Gamma$. In the main text, we take alternative routes using physical arguments in order to find the fixed point distribution, instead of directly solving this full master equation.

\end{widetext}

\section{{\normalsize{}Wavefunction and IPR from 
RG\label{sec:Wavefunction-and-IPR}}}

\begin{figure}

\includegraphics[width=\columnwidth]{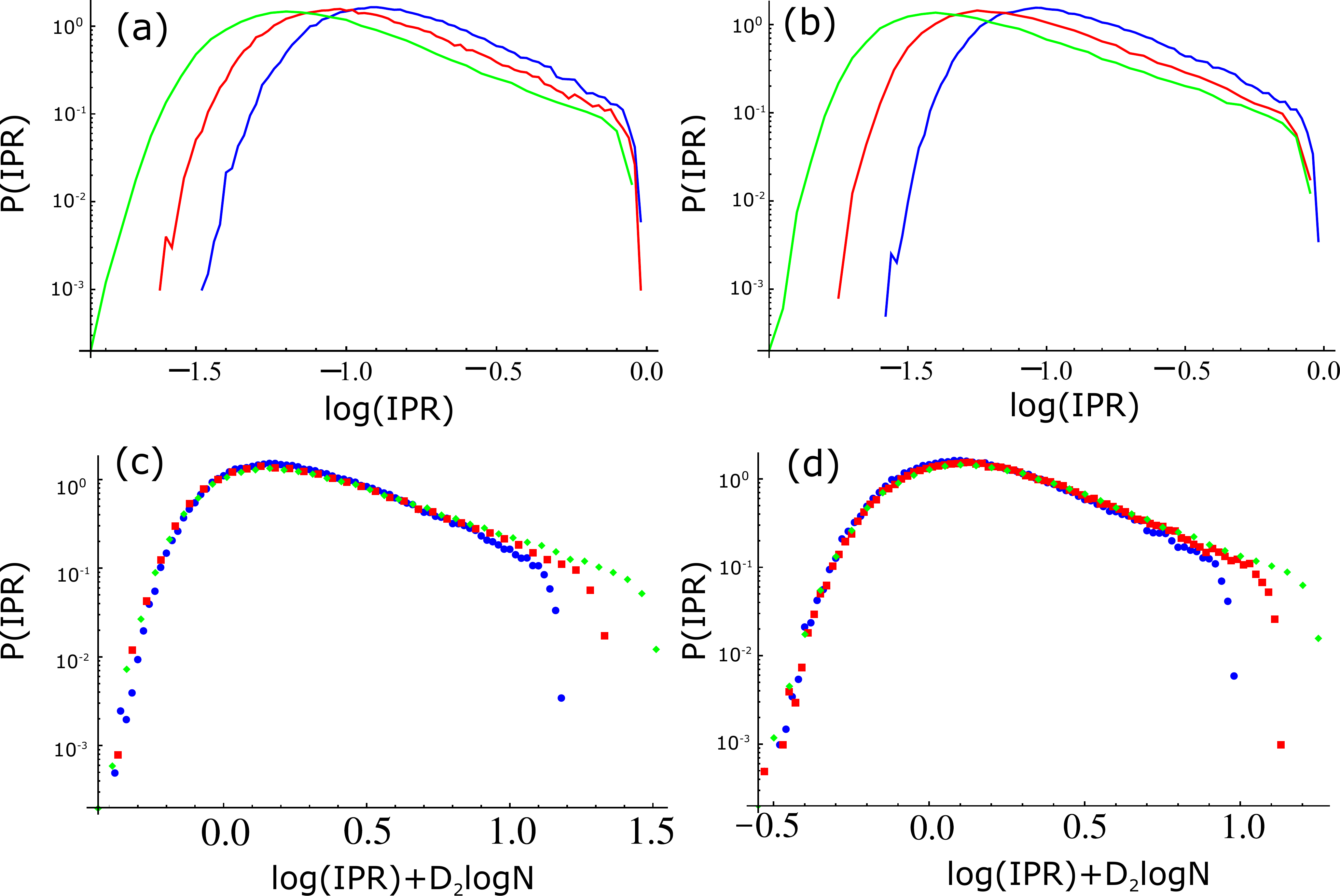}

\caption{Finite size scaling of the IPR for different system sizes $N=100$
(blue), $N=200$ (red) and $N=400$ (green). The finite size dependence and scaling collapse are shown in (a) and (c) for the
proposed RG procedure, and (b) and (d) for exact diagonalization respectively. The data are taken
for the critical point, $\alpha=1$ The initial distributions of $G_{i}^{j}$ and $h_{i}$ are Gaussian.  The fractal dimensions are
found from the scaling $\log 
\mbox{IPR}\rightarrow\log \mbox{IPR}+D_{2}\log N$. We find $D_{2}=0.5$ for the RG case (a), and $D_{2}=0.6$ for the case of
exact diagonalization (b).\label{fig:IPRRGvsEDcomparison}}
\end{figure}

In this appendix, we discuss the properties of the eigenfunctions obtained
from the RG procedure. The RG procedure consists of a rotation of the
basis states at each decimation step. Let the set of eigenfunctions
of the Hamiltonian be $\psi_{E}\left(i\right)$, where $i$ denotes
the site index and $E$ the eigenfunction label. We define a function
for the intermediate RG steps, $\psi_{E}^{m}\left(i\right)$, where
$m$ denotes the decimation step. The initial condition before any
decimation steps is $\psi_{E}^{0}\left(i\right)=\delta_{i,E}$, that is, completely localized in position space.
The eigenfunctions from the RG procedure are obtained at the end of
all the steps. We call these final functions $\psi_{E}^{F}\left(i\right)$.
Now, consider at a particular decimation step, $m$, where the bond between
sites $\left(i,j\right)$ is decimated. As discussed in Section 
\ref{sec:Flow-RG},
the corresponding rotation angle, 
$\alpha_{ij}$ was defined in Eq. (\ref{eq:maxrbondangle}).
In this step, all the intermediate functions $\psi_{E}^{m}\left(i\right)$
are modified according to
\begin{eqnarray}
\psi_{E}^{m+1}\left(i\right) & = & 
\cos\left(\alpha_{ij}\right)\psi_{E}^{m}\left(i\right)+\sin\left(\alpha_{ij}
\right)\psi_{E}^{m}\left(j\right),\label{eq:eigenfunction_evo_1}\\
\psi_{E}^{m+1}\left(j\right) & = & 
-\sin\left(\alpha_{ij}\right)\psi_{E}^{m}\left(i\right)+\cos\left(\alpha_{ij}
\right)\psi_{E}^{m}\left(j\right).\label{eq:eigenfunction_evo_2}
\end{eqnarray}
From this procedure, we can obtain the value of the IPR, by collecting the final wave functions and computing
\begin{eqnarray}
\mbox{IPR}=\sum_{i,E}\left|\psi_{E}^{F}\left(i\right)\right|^{4}
\end{eqnarray}
 The IPR scales as 
$ L^{-D_{2}}$
where $D_{2}=0$ for localized states and $D_{2}=d$ (where $d$ is the system dimension) for the extended
states \cite{EversMirlinPRL2000}. For critical states \cite{Shklovskii1993,EversMirlinPRL2000}, $0<D_{2}<d$. Furthermore, at the critical
point, the IPR distribution is postulated to
only shift and not change shape as a function of system size. We can
obtain a scaling collapse by making the appropriate rescaling, $\log 
\mbox{IPR}\rightarrow\log \mbox{IPR}+D_{2}\log N$. We plot the comparison of the IPR, at $\alpha=1$,
obtained from the RG procedure and exact diagonalization in the Fig.~\ref{fig:IPRRGvsEDcomparison}, with the scaling collapses as insets. The IPR obtained from the RG reproduces the critical behavior, with $D_{2}=0.5$ for the RG case and $D_{2}=0.6$ for exact diagonalization.

%\begin{figure}
%\includegraphics[width=1\linewidth]{OBLpaperplots/IPR_exact}

%\caption{Finite size scaling of the IPR for different system sizes $N=100$
%(blue), $N=200$ (red) and $N=400$ (green), obtained from (a) the
%proposed RG procedure and (b) exact diagonalization. The data is taken
%for the critical point, $\alpha=1$ The initial distributions of couplings is Gaussian. The fractal dimensions are
%found from the scaling collapse (insets). We find $D_{2}=0.5$ for the RG case (a), and $D_{2}=0.6$ for the case of
%exact diagonalization (b).\label{fig:IPRRGvsEDcomparison}}
%\end{figure}

\bibliographystyle{apsrev4-1}

\end{document}